\RequirePackage{fix-cm}
\documentclass[twocolumn]{svjour3}          
\smartqed  
\usepackage{graphicx}
\usepackage{graphicx, color}
\usepackage{dcolumn}

\usepackage[utf8]{inputenc}
\usepackage[T1]{fontenc}
\usepackage{mathptmx}
\usepackage{etoolbox, comment,hyperref, amsmath}
\usepackage{soul}
\usepackage{lineno}
\usepackage{algorithm2e}
\usepackage{amsfonts}
\usepackage{bm}
\def\bx{{\boldsymbol{x}}}

\def\bu{{\boldsymbol{u}}}
\def\bU{{\boldsymbol{U}}}
\def\bv{{\mathbf{v}}}
\def\by{{\mathbf{y}}}
\def\rt{{\rm{t}}}

\def\bV{{\boldsymbol{V}}}
\def\bG{{\boldsymbol{G}}}
\def\bM{{\boldsymbol{M}}}
\def\bW{{\mathbf{W}}}
\def\bD{{\mathbf{D}}}
\def\bC{{\boldsymbol{C}}}
\def\bE{{\boldsymbol{E}}}
\def\bJ{{\boldsymbol{J}}}
\def\mbJ{{\mathbf{J}}}
\def\mGamma{{\boldsymbol{\Gamma}}}
\def\bQ{{\boldsymbol{Q}}}
\def\bR{{\boldsymbol{R}}}
\def\br{{\boldsymbol{r}}}

\usepackage{xcolor}

\journalname{NoDy}
\begin{document}
	
	\title{Stability analysis of chaotic systems from data}
	
	
	\author{Georgios Margazoglou \and
		Luca Magri
	}
	
	
	\institute{G. Margazoglou \at
		Imperial College London, Aeronautics Department, South Kensington Campus
		London SW7 2AZ, United Kingdom \\
		\email{g.margazoglou@imperial.ac.uk}           
		\and
		L. Magri\at
		Imperial College London, Aeronautics Department, South Kensington Campus
		London SW7 2AZ, United Kingdom 
		\and The Alan Turing Institute,  96 Euston Road,  NW1 2DB, London, United Kingdom\\
		\email{l.magri@imperial.ac.uk}
	}
	
	\date{Published version of 10 February 2023 \href{https://doi.org/10.1007/s11071-023-08285-1}{https://doi.org/10.1007/s11071-023-08285-1}}

	\maketitle
	
	\begin{abstract}
		The prediction of the temporal dynamics of chaotic systems is challenging because infinitesimal perturbations grow exponentially. The analysis of the dynamics of infinitesimal perturbations is the subject of stability analysis. In stability analysis, we linearize the equations of the dynamical system around a reference point,  and compute the properties of the tangent space (i.e., the Jacobian). The main goal of this paper is to propose a method that infers the Jacobian, thus, the stability properties, from observables (data). 
		First, we propose the Echo State Network (ESN) with the Recycle Validation as a tool to accurately infer the chaotic dynamics from data.  
		Second, we mathematically derive the Jacobian of the Echo State Network, which provides the evolution of infinitesimal perturbations. 
		Third, we analyse the stability properties of the Jacobian inferred from the ESN, and compare them with the benchmark results obtained by linearizing the equations. The ESN correctly infers the nonlinear solution, and its tangent space properties with negligible numerical errors. In detail, we compare
		(i) the long-term statistics of the chaotic state;
		(ii) the covariant Lyapunov vectors; 
		(iii) the Lyapunov spectrum;
		(iv) the finite-time Lyapunov exponents;  
		(v) and the angles between the stable, neutral, and unstable splittings of the tangent space (the degree of hyperbolicity of the attractor).  
		This work opens up new opportunities for the computation of stability properties of nonlinear systems from data, instead of equations. 
	\end{abstract}
	\section{Introduction}\label{sec:Intro}

	Chaotic behavior has been observed and extensively studied in diverse scientific fields, initially in meteorology\cite{Lorenz63} and later in physics\cite{Ott2002,Papaphilippou2014}, chemistry, biology and engineering\cite{Strogatz2018} to name a few. Chaos appears from deterministic nonlinear equations in the form of sensitivity to initial conditions, aperiodic behavior, and short predictability. A successful  mathematical tool for the analysis of chaos is provided by stability analysis. By applying infinitesimal perturbations to a system's trajectory, we can classify its stability along different directions, and compute the properties of its linear tangent space. 
	
	Stability analysis relies on the linearization of the dynamical equations, which requires the Jacobian of the system. The key quantities that characterize chaotic dynamics, and many other related physical properties, such as dynamical entropies and fractal dimensions, are the Lyapunov Exponents (LEs)\cite{Ruelle1980,Eckmann_Ruelle1985}, which are the  eigenvalues of the Oseledets matrix\cite{Oseledets1968}. There are several numerical methods to extract the LEs based on the Gram-Schmidt orthogonalization procedure\cite{Benettin1980,Shimada_Nagashima1979,Eckmann_Ruelle1985}. The relevant eigenvectors are the corresponding Lyapunov vectors that constitute a coordinate dependent orthogonal basis of the linear tangent space. 
	Instead, an intrinsic and norm-independent basis, which is also time invariant and covariant with the dynamics is given by the Covariant Lyapunov vectors (CLVs). Crucially, CLVs are able to provide information on the local structure of chaotic attractors\cite{Ginelli2013}. This viewpoint allows the study of an attractor's topology with the occurrence of critical transitions\cite{Schubert2015,Vannitsem2016,Sharafi2017,Brugnago2020}, paving the way for CLVs to be considered as precursors to such phenomena.  
	
	The previous exposition is traditionally related with model-based approaches, as it relies on the knowledge of a system's dynamical equations. However, studying the stability properties of observed data, where equations are not necessarily known, is hard; there are few approaches e.g.~\cite{Wolf1985,Rosenstein1993}, relying on the delayed coordinates attractor reconstruction by Takens\cite{Takens1981}. The recent breakthrough of data-driven (model-free) approaches poses the reasonable question: Can we use the rich knowledge of dynamical systems theory for model-free approaches?  Indeed, although at early steps, the use of advanced Machine Learning (ML) techniques for complex systems has shown promising potential in applications ranging from weather and climate prediction and classification \cite{Rasp2018,Dueben2018,Margazoglou2021} to fluid flows prediction and optimization \cite{Verma2018,Kochkov2021,Huhn_Magri_2022}, among others. The overarching goal of this work is to propose a machine learning approach to accurately learn and infer the ergodic properties of prototypical chaotic attractors, and in particular to extract LEs and CLVs from data.

	The Recurrent Neural Networks (RNNs) constitute a promising type of ML to address chaotic behavior. Thanks to their architecture, the RNNs are suitable for processing sequential data, typically encountered in speech and language recognition, or time-series prediction\cite{Goodfellow2016}. In particular, they are proven to be universal approximators\cite{Schafer2006,Grigoryeva2018}, and are able to capture long-term temporal patterns, i.e.~they possess memory. A key piece of their architecture is that they maintain a hidden state that evolves dynamically, effectively allowing the RNNs to be treated as dynamical systems, and in particular as discrete neural differential equations\cite{Chen2018}. Thus, RNNs lend themselves to being analysed with  dynamical systems theory, allowing the study of stability properties from the dynamics they have learned. By exploiting this here, we derive the RNN's Jacobian and infer the linear dynamics from data. 
	
	Recently there have been significant advancements in employing RNNs to learn chaotic dynamics \cite{Lu2018,Pathakchaos2017,PathakPRL2018,Vlachas2018,Vlachas2020,Borra2020,Doan2020,Doan2021_prsa,Racca2021}, where two core objectives are studied: (1) the time-accurate prediction of chaotic fluctuations and maximization of the prediction horizon; and (2) accurately learning the ergodic properties of chaotic attractors.  The first objective has been addressed by one of the co-authors in \cite{Doan2020,Doan2021_prsa,Racca2021} for several prototypical chaotic dynamical systems using the same RNN architecture as the present work. Here we address the second objective by extending the recent works\cite{Pathakchaos2017,PathakPRL2018,Vlachas2020}, where the LEs of the Lorenz 63\cite{Lorenz63} and the one-dimensional Kuramoto-Sivashinsky equation \cite{Kuramoto1978} were retrieved from trained RNNs.
	
	In this work we employ a specific architecture of the RNN, a type of reservoir computer, the Echo State Network (ESN)\cite{Lukosevicius2012} and train it with a diverse set of four prototypical chaotic attractors. The objective of this paper is twofold; first the accurate learning and inference of the ergodic properties of the chaotic attractors by the ESN. This is accomplished by thoroughly comparing the long-term statistics of (i) degrees of freedom, (ii) LEs, (iii) finite-time LEs, and (iv) angles of the CLVs. Second, by comparing the distribution of (i) finite-time LEs, and (ii) angles of CLVs on the topology of the attractor, providing a strict test of the ESN's capability to accurately learn intrinsic chaotic properties.

	The paper is organized as follows. Section \ref{sec:theory} presents the necessary tools for our study. In particular, Sect.~\ref{sec:CLVs} provides a brief introduction to the relevant concepts and quantities from dynamical systems, such as LEs and CLVs. Then Sect.~\ref{sec:ESN} describes the architecture of the ESN, while Sect.~\ref{sec:val} its validation strategies. Section \ref{sec:results} presents our main results, which are divided into two subsections; Sect.~\ref{sec:low_dyn_syst} devoted in low-dimensional systems, namely the Lorenz 63 \cite{Lorenz63} and R\"ossler\cite{Rossler1976} attractors; and Sect.~\ref{sec:high_dyn_syst} showing results on the Charney-DeVore\cite{CdV1979} and the Lorenz 96\cite{Lorenz96} attractors. Finally, we summarize our results and provide future perspectives in the conclusions in Sect.~\ref{sec:conc}. The appendix~\ref{app:1} presents the two algorithms to extract the LEs and CLVs from the ESN. Additionally, appendix ~\ref{app:2} provides further tests on the robustness of our methodology.

	\section{Background}\label{sec:theory}
	In the following two subsections, we summarize the key theory that underpins the stability of chaotic systems (Sect.~\ref{sec:CLVs}) and reservoir computers (Sect.~\ref{sec:ESN}). 
	
	\subsection{Stability of chaotic systems}\label{sec:CLVs}
	
	We consider a state $\bx(t) \in \mathbb{R}^D$ with $D$ degrees of freedom, which is governed by a set of nonlinear ordinary differential equations
	\begin{equation}
		\frac{d\bx}{dt} = f(\bx),
		\label{eq:dynamicalsyst}
	\end{equation}
	where $f(\bx) : \mathbb{R}^D \to \mathbb{R}^D$ is a smooth nonlinear function. 
	Equation~\eqref{eq:dynamicalsyst} defines an autonomous dynamical system. Hence, the dynamical system exists in a phase space of dimension $D$, equipped with a suitable metric, and is associated with a certain measure $\mu$ that we assume to be preserved (invariant).  To investigate the stability  of the dynamical system \eqref{eq:dynamicalsyst}, we perturb the state by first-order perturbations as 
	\begin{align}
		&\bx+\bu, \label{eq:firstor}
		&\bx\sim\mathcal{O}(1), \;\;\;\bu\sim\mathcal{O}(\epsilon), \;\;\;\epsilon\to 0. 
	\end{align}
	
	By substituting  decomposition~\eqref{eq:firstor} into \eqref{eq:dynamicalsyst} and collecting the first-order terms $\sim\mathcal{O}(\epsilon)$, we obtain the governing equation for the first-order perturbations (i.e., linear dynamics)
	\begin{equation}
		\frac{d\bu}{dt} = \bJ(\bx(t))\bu,
		\label{eq:tangdynamics}
	\end{equation}
	where $J_{ij}=\frac{\partial f_i(x)}{\partial x_j} \in \mathbb{R}^{D\times D}$ are the components of the Jacobian, $\bJ(\bx(t))$, which is in general time-dependent. The perturbations $\bu$ evolve on the linear tangent space at each point $\bx(t)$. 
	The goal of stability analysis is to compute the growth rate of infinitesimal perturbations, which is achieved by computing  the Lyapunov exponents and a basis of the tangent space with the Covariant Lyapunov Vectors. To do so, we numerically time-march $K\leq D$ tangent vectors, $\bu_i\in \mathbb{R}^D$, as columns of the matrix $\bU\in \mathbb{R}^{D\times K}$, $\bU = [\bu_1, \bu_2,\dots,\bu_K]$
	\begin{equation}
		\frac{d\bU}{dt} = \bJ(\bx(t))\bU.
		\label{eq:Qtangdynamics}
	\end{equation}
	
	Geometrically, Eq.~\eqref{eq:Qtangdynamics} describes the tangent space around the state $\bx(t)$. Starting from $\bx(t=t_0)=\bx_0$ and $\bU(t_0)=\mathbb{I}$, Eqs.~\eqref{eq:dynamicalsyst} and \eqref{eq:Qtangdynamics} are numerically solved with a time integrator. 
	As explained in the subsequent paragraphs, in a chaotic system, almost all nearby trajectories diverge exponentially fast with an average rate equal to the leading Lyapunov exponent. Hence, the tangent vectors  align exponentially fast with the leading Lyapunov vector, $\bu_1$. (`Almost all' means that the set of perturbations that do not grow with the largest Lyapunov exponents has a zero measure.) 
	To circumvent this numerical issue, it is necessary to periodically orthonormalize the tangent space basis during time evolution, using a QR-decomposition of $\bU$, as  $\bU(t)=\bQ(t) \bR(t,\Delta t)$ (see \cite{Benettin1980,Shimada_Nagashima1979}) and updating the columns of $\bU$ with the columns of $\bQ$, i.e.~$\bU \leftarrow \bQ$. 
	The matrix $\bR(t,\Delta t)\in \mathbb{R}^{K\times K}$ is upper-triangular and its diagonal elements $[\bR]_{i,i}$ are the local growth rates over a time span $\Delta t$ of the (now) orthonormal vectors $\bU$, which are also known as backward Gram-Schmidt vectors (GSVs)\cite{Ginelli2007,Ginelli2013}. The Lyapunov spectrum is given by \footnote{ The Oseledets' theorem~\cite{Oseledets1968,Eckmann_Ruelle1985,Ginelli2013} establishes the existence of Lyapunov exponents (LEs) for a generic set of orbits under fairly general assumptions. In particular, the Oseledets' theorem enables the extension of Lyapunov stability analysis to any trajectory of a dynamical system defined on a Riemannian manifold of dimension $N$ and equipped with a suitable metric, including fixed points and periodic orbits.}
\begin{equation}
\lambda_i = \lim\limits_{T\to\infty}\frac{1}{T} \int_{t_0}^{T}\ln [\bR(t,\Delta t)]_{i,i}dt.
\label{eq:LEs}
\end{equation}

The algorithm \ref{alg:LEs} in the appendix~\ref{app:1} is a pseudocode for the calculation of the LEs for the ESN following \cite{Pathakchaos2017,Vlachas2020}. The sign of the Lyapunov exponents indicates the type of the attractor. If the leading exponent is negative, $\lambda_1<0$, the attractor is a fixed point. If $\lambda_1=0$, and the remaining exponents are negative, the attractor is a periodic orbit. If at least a Lyapunov exponent is positive, $\lambda_1>0$, the attractor is chaotic. In chaotic systems, the Lyapunov time $\tau_\lambda = \frac{1}{\lambda_{1}}$ defines a characteristic timescale for two nearby orbits to separate, which gives a scale  of the system's predictability horizon\cite{Boffetta2002}.

The GSVs, $\bU$, constitute a norm-dependent orthonormal basis,  which is not time-reversible, due to the frequent orthogonalizations via the QR decomposition.   Instead, the Covariant Lyapunov Vectors (CLVs) $\bV = [\bv_1, \bv_2,\dots,\bv_K]$ (each CLV $\bv_i\in \mathbb{R}^D$ is a column of $\bV$) form a norm-independent and time-invariant basis of the tangent space, which is covariant with the dynamics. The latter features of the CLVs, which are not possessed  by the GSVs, allow us to examine individual expanding and contracting directions of a given dynamical system, thus providing an intrinsic geometrical interpretation of the attractor\cite{Eckmann_Ruelle1985,Ginelli2013}, as well as a hierarchical decomposition of spatiotemporal chaos, thanks to their generic localization in physical space\cite{Ginelli2007}. Each bounded non-zero CLV, i.e.~$0<||\bv_i||<\infty$, satisfies the following equation
\begin{equation}
\frac{d\bv_i}{dt} = \bJ(\bx(t))\bv_i - \lambda_i\bv_i,
\label{eq:clv_eq}
\end{equation}
which shows that the CLV is evolved by the tangent dynamics $\bJ(\bx(t))\bv_i$, while the extra term $- \lambda_i\bv_i$ guarantees that its norm is bounded\cite{Huhn_Magri_2020}. The name ``covariant'' means that the $i$th CLV at time $t_1$, $\bv_i(\bx(t_1))$, maps at $\bv_i(\bx(t_2))$ at time $t_2$, and vice versa. Mathematically, if $\bM(t,\Delta t)=\exp(\int_{t}^{t+\Delta t} \bJ(\bx,\tau)d\tau)$ is the system's tangent evolution operator (which contains a path-ordered exponential), covariant means  $\bM(t, \Delta t) \bv_i(t) = \bv_i(t+\Delta t)$; time-invariance  of CLVs naturally arises from the previous expression, as $\bM(t, -\Delta t) \bv_i(t+\Delta t) = \bv_i(t)$. If the Lyapunov spectrum is non-degenerate (such as for the cases considered here) each CLV $\bv_i$ is associated with the Lyapunov exponent $\lambda_i$, and is uniquely defined (up to a phase).

An important subclass of chaotic systems are uniformly hyperbolic systems, which have a uniform splitting between expanding and contracting directions, i.e., there are no tangencies between the unstable, neutral, and stable subspaces \cite{Ruelle1997} that form the tangent space. Because of their simple geometrical structure many theoretical tools have been developed in recent years. Hyperbolic systems have structurally stable dynamics, and linear response, meaning that their statistics vary smoothly with parameter variations\cite{Lucarini2014}. In practice, violations of hyperbolicity are commonly reported in the literature\cite{Blonigan2017,Wormell2022,Huhn_Magri_2020}, whereas true hyperbolic systems are rare\cite{Kuznetsov2012}. 
Thanks to the chaotic hypothesis \cite{Ruelle1980,Gallavotti_Coen1995_PRL,Gallavotti_Coen1995_JSP},  high-dimensional chaotic systems  can be practically treated as hyperbolic systems, i.e.~using techniques developed for hyperbolic systems, regardless of hyperbolicity violations. This is because many convenient statistical properties of uniformly hyperbolic systems, such as ergodicity, existence of physical invariant measures, exponential mixing and well-defined time averages with large deviation laws \cite{LebowitzSpohn1999,Lepri1998}, can be found in the macroscopic scale dynamics of certain large non-uniformly hyperbolic systems\cite{Lucarini2014}. 

An application of CLVs is to assess the degree of hyperbolicity of the underlying chaotic dynamics. The tangent space  of hyperbolic systems, at each point $\bx$,  can be directly decomposed into three invariant subspaces,  $\bE^U_\bx \oplus \bE^N_\bx \oplus \bE^S_\bx $. Here $\bE^U_\bx$ is the unstable subspace composed by the CLVs associated with positive LEs, $\bE^N_\bx$ is the neutral subspace spanned by the CLVs associated with the zero LEs, and  $\bE^S_\bx$ is the stable subspace spanned by the CLVs associated with negative LEs. In hyperbolic systems, the distribution of angles between subspaces is bounded away from zero. In Sect.~\ref{sec:results} we will study in detail the angles $\theta_{U,N}$, $\theta_{U,S}$, and $\theta_{N,S}$  between pairs of the subspaces, and compare the ability of the ESN to accurately learn both the long-term statistics, and the phase space finite-time variability of the angles. Because the GSVs are mutually orthogonal, they cannot assess the degree of hyperbolicity of the attractor. Moreover, CLVs are key to the optimization of chaotic acoustic oscillations \cite{Huhn_Magri_2020}, as well as in reduced-order modelling\cite{Yang2009}; they can reveal two uncoupled subspaces of the tangent space, one that comprises the physical modes carrying the relevant information of the trajectory, and another composed of strongly decaying spurious modes\cite{Ginelli2013}. Two recent attempts to extract CLVs from data driven approaches, which do not employ a neural network, can be found in\cite{Viennet2022,Christoph2022}.

We explain the algorithm we employ to compute the CLVs; for further details we refer the interested reader to \cite{Ginelli2007,Ginelli2013,Huhn_Magri_2020}.
The GSVs are generated by numerically solving Eqs.~\eqref{eq:dynamicalsyst} and \eqref{eq:Qtangdynamics} simultaneously, and performing a QR-decomposition every $m$ timesteps. In this way, after a time-lapse $\Delta t$, the GSVs at time $t+\Delta t$ are given by
\begin{equation}
\bM(t,\Delta t) \bU(t) = \bU(t+ \Delta t) \bR(t, \Delta t).
\label{eq:gsv}
\end{equation}
We can define the CLVs $\bV(t)$ in terms of the GSVs as 
\begin{equation}
\bV(t) = \bU(t) \bC(t),
\label{eq:gsv2clv}
\end{equation}
where $\bC$ is an upper triangular matrix that contains the CLV expansion coefficients, $[\bC(t)]_{ji} = c^{j,i}(t)$, for $j\leq i$. Hence, the objective is to calculate $\bC(t)$. Because the CLVs have by choice a unit norm, each column of the matrix $\bC$ has to be normalized independently, i.e.~$\sum_{j=1}^{i} (c_{j,i}(t))^2 = 1, \forall i$.

We start by writing the evolution equation of the CLVs as 
\begin{equation}
\bM(t,\Delta t) \bV(t) = \bV(t+ \Delta t) \bD(t, \Delta t).
\label{eq:clv}
\end{equation}
We can re-write Eq.~\eqref{eq:clv} via Eq.~\eqref{eq:gsv2clv}
\begin{eqnarray}
\bU(t+ \Delta t) \bC(t+ \Delta t) \bD(t, \Delta t) &=& \bM(t,\Delta t) \bU(t) \bC(t)\\
&=& \bU(t+ \Delta t) \bR(t, \Delta t) \bC(t), \nonumber
\label{eq:clvderiv}
\end{eqnarray}
and solve with respect to $\bC(t)$
\begin{eqnarray}
\bC(t) = \bR^{-1}(t, \Delta t) \bC(t+ \Delta t) \bD(t, \Delta t).
\label{eq:C}
\end{eqnarray}
This equation is evolved backwards in time starting from the end of the forward-in-time simulation. We employ the \texttt{solve\_triangular} routine of scipy\cite{SciPy2020} to invert $\bR(t,\Delta t)$ and solve with respect to $\bC(t)$. The $\bC$ and $\bD$ matrices are initialized to the identity matrix $\mathbb{I}$. We leave a sufficient spin-up and spin-down transient time at the beginning and end of our total time window, before we compute the CLVs via Eq.~\eqref{eq:gsv2clv}, to ensure that they are  converged. The algorithm \ref{alg:CLVs} in the appendix~\ref{app:1} is a pseudocode for the calculation of the CLVs.

To estimate the expansions and contractions of the tangent space on ﬁnite-time intervals of length $\Delta t = t_2 - t_1$, we compute the finite time Lyapunov exponents (FTLEs)  as $\Lambda_i = \frac{1}{\Delta t} \ln [\bR]_{i,i}$. Hence, $\lambda_i$ is  the long-time average of $\Lambda_i$. The FTLE $\Lambda_1$ physically quantifies the exponential growth rate of a vector $\bu_1$ during the time interval $\Delta t$, therefore, $\Lambda_2$ quantifies the exponential growth rate of the vector $\bu_2$ that is orthogonal to $\bu_1$ by construction. Hence, as the GSVs form an orthogonal basis, looking at individual FTLEs for $\Lambda_i$, $i\ge2$, lacks a physical meaning. Instead, the sum of the ﬁrst $n$ FTLEs is a growth rate in $\Delta t$ for a typical $n$-dimensional volume ${\rm Vol_n}$ in the tangent space \cite{Shimada_Nagashima1979,kuptsov2018}
\begin{eqnarray}
\sum_{i=1}^{n} \Lambda_i &=& \frac{1}{\Delta t} \sum_{i=1}^{n} \ln[\bR(t,\Delta t)]_{i,i} = \frac{1}{\Delta t}  \ln \prod_{i=1}^{n}[\bR(t,\Delta t)]_{i,i} \nonumber \\
&=& \frac{1}{\Delta t}  \ln{{\rm Vol_n}(\Delta t)}.
\end{eqnarray}
Accordingly, the diagonal matrix $\bD(t, \Delta t)$ contains the CLV local growth factors of $\gamma_i(t, \Delta t) = ||\bM(t, \Delta t) \bv_i(t)||$, i.e.~$[\bD(t, \Delta t)]_{i,j}=\delta_{i,j} \gamma_i(t, \Delta t)$. We can extract the finite-time Covariant Lyapunov exponents (FTCLEs) from the logarithm of these growth factors for a time interval $\Delta t$
\begin{equation}
\Lambda^c_i = \frac{1}{\Delta t} \ln [\bD]_{ii}.
\label{eq:ftcles}
\end{equation}
Each FTCLE quantifies a finite-time exponential expansion or contraction rate along a covariant direction given by $\bv_i$. Hence each individual FTCLE has a physical interpretation, in contrast to the FTLEs, as explained before. On the other hand, now the sums of FTCLEs lack a physical meaning\cite{kuptsov2018}. The long-time average of the FTCLEs is equal to the Lyapunov exponents, $\lambda_i = \lim\limits_{T\to\infty}\frac{1}{T} \int_{t_0}^{T}\Lambda^c_i(t)dt$.

\subsection{Echo State Network}\label{sec:ESN}

\begin{figure}[t]
\centering
\includegraphics[width=.4\textwidth]{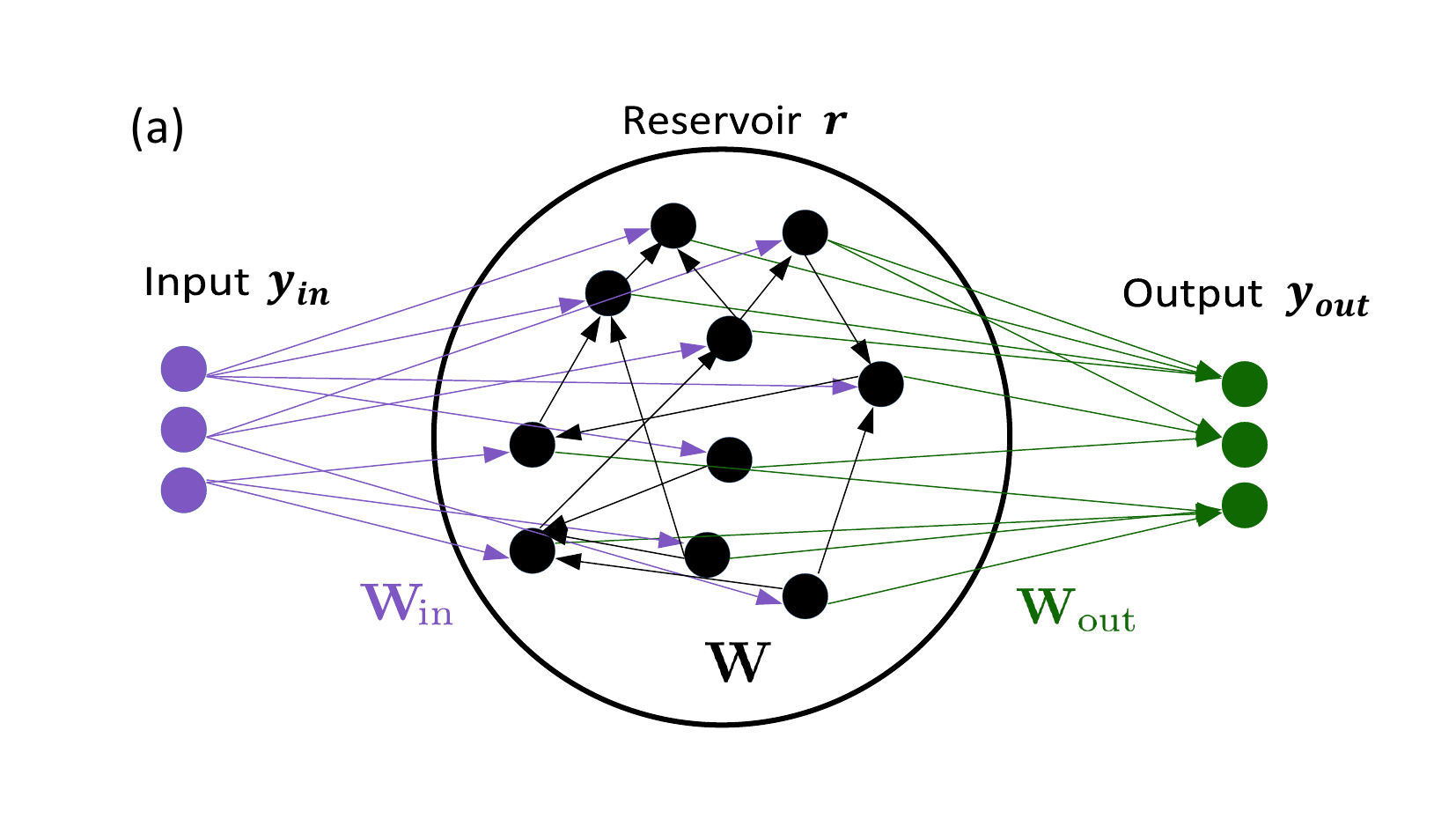}
\includegraphics[width=0.35\textwidth]{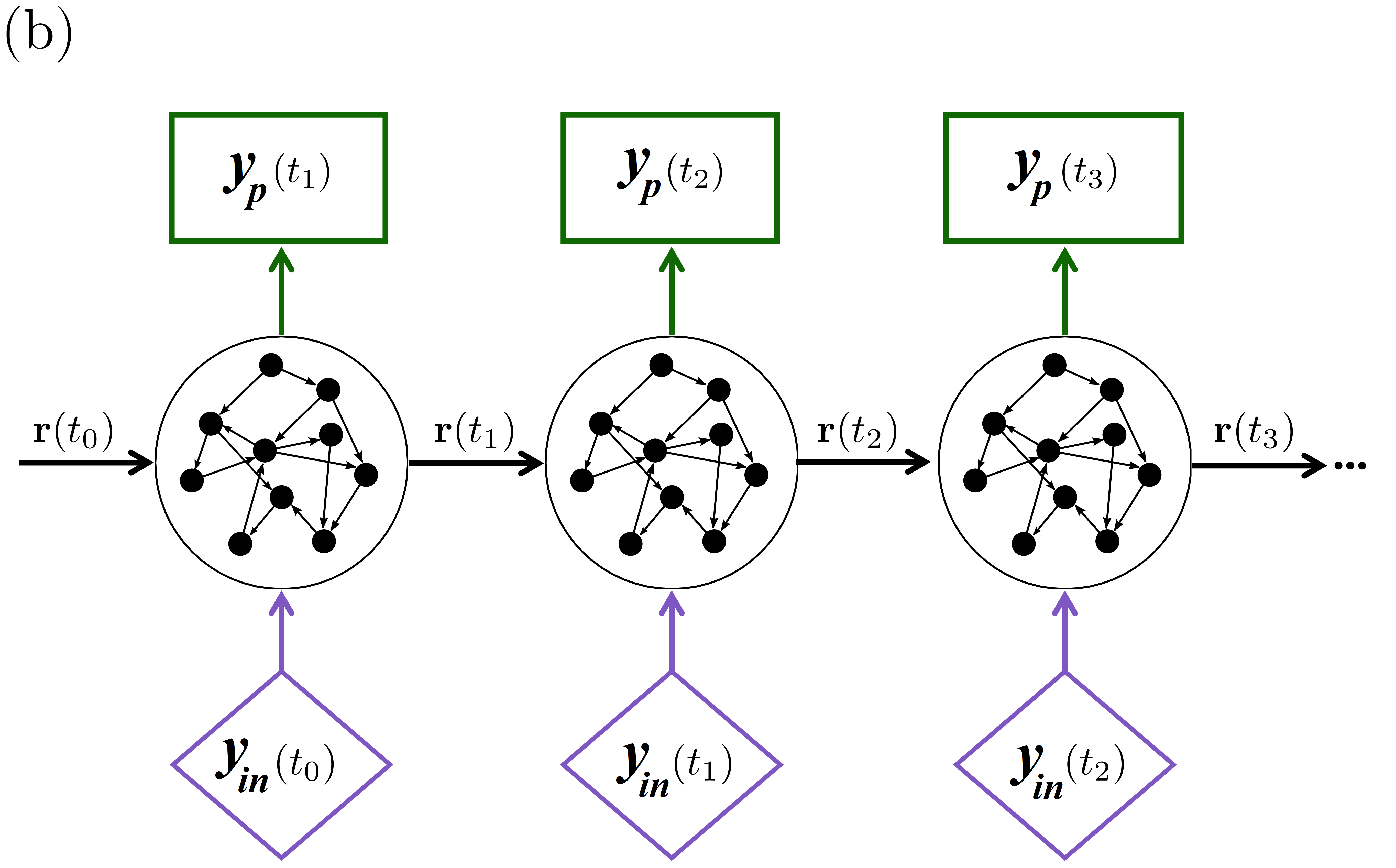}
\includegraphics[width=0.35\textwidth]{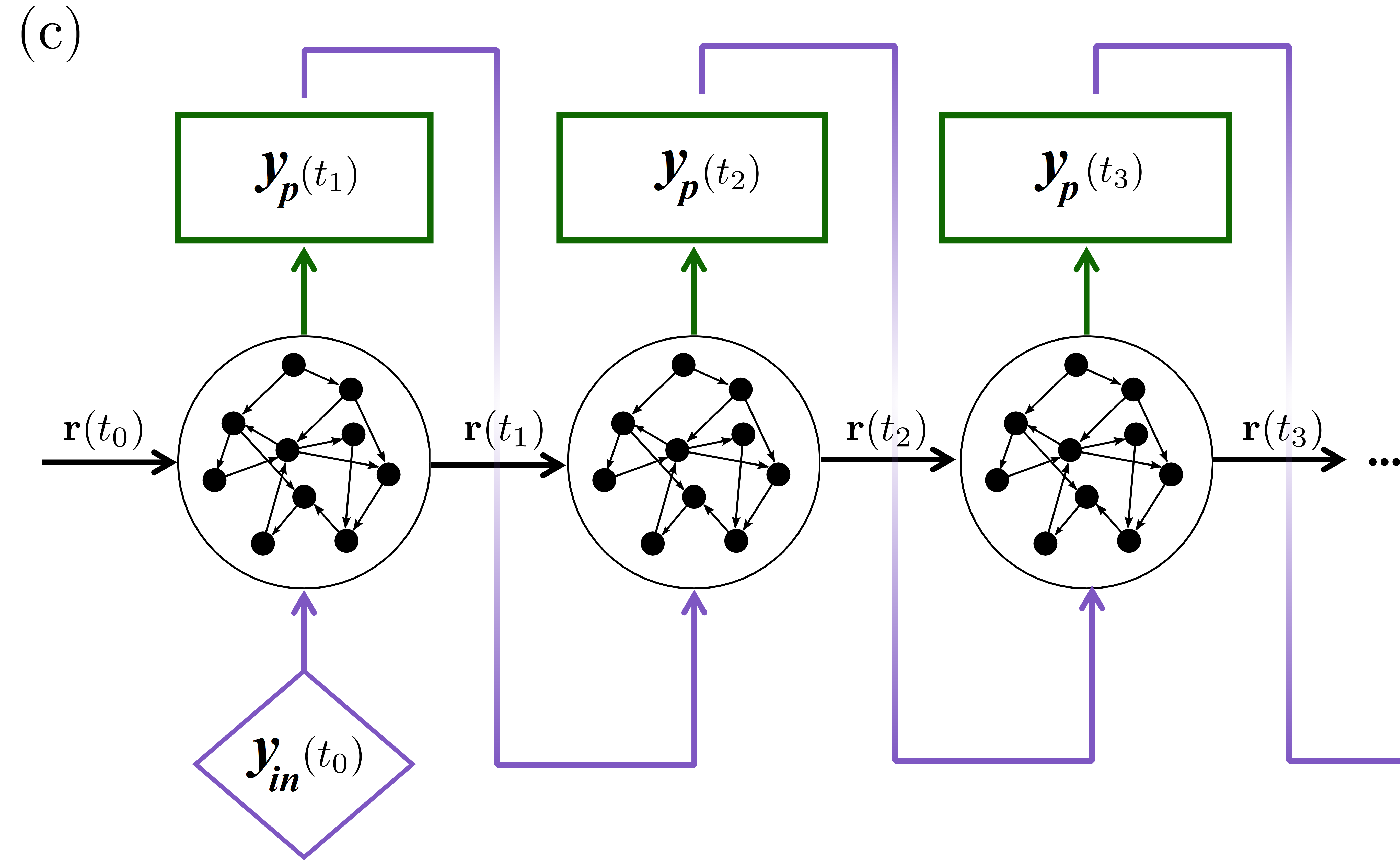}
\caption{(a) Schematic representation of the echo state network. (b) Open-loop and (c) closed-loop  configurations.}
\label{fig:ESN}
\end{figure}

The solution of a dynamical system is a time series. 
From a data analysis point of view, a time series is a sequentially ordered set of values, in which the order is provided by time. In a discrete setting, time can be thought of as an ordering index. For sequential data, and hence time series, recurrent neural networks (RNNs) are designed to infer the temporal dynamics through their internal hidden state. However, training  RNNs, such as  Long-Short term memory (LSTM)\cite{Hochreiter1997} networks and Gated Recurrent Units (GRUs)\cite{Cho2014}, requires backpropagation through time, which can be a demanding computational task due to the long-lasting time dependencies of the hidden states \cite{Werbos1990}. 
This issue is overcome by Echo State Networks (ESNs)\cite{Jaeger2004,Lukosevicius2012}, a RNN that is a type of reservoir computer, of which the recurrent weights of the hidden state (commonly named ``reservoir'') are randomly assigned and possess low connectivity. Therefore, only the hidden-to-output weights are trained leading to a simple quadratic optimization problem, which does not require backpropagation (see Fig.~\ref{fig:ESN}(a) for a graphical representation). The  reservoir acts as a memory of the observed state history. ESNs have demonstrated accurate inference of chaotic dynamics, such as in\cite{Lu2018,Pathakchaos2017,PathakPRL2018,Vlachas2018,Vlachas2020,Borra2020,Doan2020,Doan2021_prsa,Racca2021,Huhn2020proc}.

An Echo state network maps the state from time index $\rt_i$ to index $\rt_{i+1}$ as follows (with a slight abuse of notation,  the discrete time is denoted  $\rt_i$). 
The evolution equations of the reservoir state and output are governed, respectively, by \cite{Lukosevicius2012,Racca2021}
\begin{eqnarray}
\label{eq:esn_r}
\br(\rt_{i+1}) &=& \tanh\left([\hat{\by}_{\mathrm{in}}(\rt_{i});b_\mathrm{in}]^T\mathbf{W}_{\mathrm{in}}+\br(\rt_{i})^T\mathbf{W}\right), \\ \by_{\mathrm{p}}(\rt_{i+1}) &=& [\br(\rt_{i+1});1]^T\mathbf{W}_{\mathrm{out}}; 	\label{eq:esn_y}
\end{eqnarray}
where at any discrete time $\rt_i$ the input vector, $\by_{\mathrm{in}}(\rt_{i}) \in \mathbb{R}^{N_y}$, is mapped into the reservoir state $\br \in \mathbb{R}^{N_r}$, by the input matrix, $\mathbf{W}_{\mathrm{in}} \in \mathbb{R}^{(N_y+1)\times N_r }$, where $N_r \gg N_y$. The updated reservoir state $\br(\rt_{i+1})$ is calculated at each time iteration as a function of the current input $\hat{\by}_{\mathrm{in}}(\rt_{i})$ and its previous value $\br(\rt_{i})$ via Eq.~\eqref{eq:esn_r}, and then is involved in the calculation of the predicted output,  $\by_{\mathrm{p}}(\rt_{i+1})\in \mathbb{R}^{N_y}$ via Eq.~\eqref{eq:esn_y}. Here, $\hat{(\;\;)}$ indicates normalization by the maximum-minus-minimum range of $\by_{\mathrm{in}}$ in training set, component-wise, $(^T)$ indicates matrix transposition, (;) indicates array concatenation, $\mathbf{W} \in \mathbb{R}^{N_r\times N_r}$ is the state matrix, $b_{\mathrm{in}}$ is the input bias and  $\mathbf{W}_{\mathrm{out}} \in \mathbb{R}^{(N_{r}+1)\times N_y}$ is the output matrix. In our applications, the dimension of the input and output vectors is equal to the dimension of the  physical system of Eq.~\eqref{eq:dynamicalsyst}, i.e.~$N_y \equiv D$.

The matrices $\mathbf{W}_{\mathrm{in}}$ and $\mathbf{W}$ are (pseudo)randomly generated and fixed, whilst the weights of the output matrix, $\mathbf{W}_{\mathrm{out}}$, are the only trainable elements of the network. The input matrix, $\mathbf{W}_{\mathrm{in}}$, has only one element different from zero per row, which is sampled from a uniform distribution in $[-\sigma_{\mathrm{in}},\sigma_{\mathrm{in}}]$, where $\sigma_{\mathrm{in}}$ is the input scaling. The state matrix, $\textbf{W}$, is an Erd\"os-Renyi matrix with average connectivity $d$, in which each neuron (each row of $\mathbf{W}$) has on average only $d$ connections (i.e.~non-zero elements), which are obtained by sampling from a uniform distribution in $[-1,1]$. The echo state property enforces the independence of the reservoir state on the initial conditions, which is satisfied by rescaling $\textbf{W}$ by a multiplication factor, such that the absolute value of the largest eigenvalue \cite{Lukosevicius2012}, i.e., the spectral radius, is smaller than unity. Following \cite{Pathakchaos2017,Huhn2020proc,Racca2021,Racca2022}, we add a bias in the input and output layers to break the inherent symmetry of the basic ESN architecture. Specifically, the input bias, $b_{\mathrm{in}}$ is a hyperparameter, selected in order to have the same order of magnitude as the normalized inputs, $\mathbf{\hat{y}}_{\mathrm{in}}$. Differently, the output bias is determined by training the weights of the output matrix, $\mathbf{W}_{\mathrm{out}}$.

In Figs.~\ref{fig:ESN}(b)-(c) we present the two types of configurations with which the ESN can run, i.e. in open-loop or closed-loop, respectively. Running in open-loop is necessary for the training stage, as the input data is fed at each step, allowing for the calculation of the reservoir timeseries $\br(\rt_{i})$, $\rt_{i} \in [0,T_{\mathrm{train}}]$, which need to be stored. There is an initial transient time window, the ``washout interval'', where the output $\by_{\mathrm{p}}(\rt_{i})$ is not computed. This allows for the reservoir state to satisfy the echo state property, i.e.~making it independent of the arbitrarily chosen initial condition, $\br(t_0) = {0}$, while also synchronizing it with respect to the current state of the system.

The training of the output matrix, $\mathbf{W}_{\mathrm{out}}$, is performed after the washout interval, and involves the minimization of the mean square error between the outputs and the data over the training set
\begin{equation}
\textrm{MSE} = \frac{1}{N_{\mathrm{tr}}N_y} \sum_{i=0}^{N_{\mathrm{tr}}} || \by_{\mathrm{p}}(\rt_{i}) - \by_{\mathrm{in}}(\rt_{i})||^2,
\label{eq:MSE}
\end{equation}
where $||\cdot||$ is the $L_2$ norm, $N_{\mathrm{tr}}+1$ is the total number of data in the training set, and $\by_{\mathrm{in}}$ the input data on which the ESN is trained.  Training the ESN is performed by solving with respect to $\mathbf{W}_{\mathrm{out}}$ via ridge regression of
\begin{equation}
\label{RidgeReg}
(\mathbf{R}\mathbf{R}^T + \beta \mathbb{I})\mathbf{W}_{\mathrm{out}} = \mathbf{R} \mathbf{Y}_{\mathrm{d}}^T,
\end{equation}
where $\mathbf{R}\in\mathbb{R}^{(N_r+1)\times N_{\mathrm{tr}}}$ and $\mathbf{Y}_{\mathrm{d}}\in\mathbb{R}^{N_y\times N_{\mathrm{tr}}}$ are the horizontal concatenation of the reservoir states with bias, $[\br(\rt_{i});1]$, $\rt_i \in [0,T_{\mathrm{train}}]$, and of the output data, respectively; $\mathbb{I}$ is the identity matrix and $\beta$ is the Tikhonov regularization parameter \cite{Tikhonov1995}. 

On the other hand, in the closed-loop configuration (Fig. \ref{fig:ESN}(c))  the output $\by_{\mathrm{p}}$ at time step $\rt_{i}$, is used as an input at time step $\rt_{i+1}$, in a recurrent manner, allowing for the autonomous temporal evolution of the network. The closed-loop configuration is used for validation (i.e. hyperparameter tuning, see Sec.\ref{sec:val}) and testing, but not for training. For our purposes, we independently train $N_{\mathrm{ESN}}\in[5,10]$ networks, of which we take the ensemble average to increase the statistical accuracy of the prediction and evaluate its uncertainty. We  start with $N_{\mathrm{ESN}}=10$ trained networks, but during post-processing we may discard any network that shows spurious temporal evolution.  The $N_{\mathrm{ESN}}$ networks are statistically independent thanks to: 1) initializing the random matrices $\mathbf{W}_{\mathrm{in}}$ and $\mathbf{W}$ with different seeds, and 2) training each network with chaotic timeseries staring from different initial points on the attractor.

\subsubsection{Jacobian of the ESN}
In this subsection, we mathematically derive the Jacobian of the Echo State Network. 
Equations \eqref{eq:esn_r}-\eqref{eq:esn_y} are a discrete map\cite{Ott2002,Vlachas2020},
\begin{eqnarray}
\br(\rt_{i+1}) &=& f(\by(\rt_{i}), \br(\rt_{i})) = \tanh\left([\hat{\by}(\rt_{i});b_\mathrm{in}]^T\mathbf{W}_{\mathrm{in}}+\br(\rt_{i})^T\mathbf{W}\right),\nonumber\\
\by(\rt_{i+1}) &=& 
[\br(\rt_{i+1});1]^T \textbf{W}_{\mathrm{out}}, \nonumber
\end{eqnarray}
and the continuous-time formulae derived for the Lyapunov exponents and CLVs in Sect.~\ref{sec:CLVs} can be adapted for a discrete-time system. The Jacobian of the ESN reservoir  is the total derivative of the hidden state dynamics at a single timestep\cite{Pathakchaos2017} 
\begin{eqnarray}
\mbJ(\br(\rt_{i+1})) &=& \frac{d\br(\rt_{i+1})}{d\br(\rt_{i})}  = \frac{df(\by(\rt_{i}), \br(\rt_{i}))}{d\br(\rt_{i})} \nonumber \\ 
&=& 
\frac{\partial f(\by(\rt_{i}), \br(\rt_{i}))}{\partial \by(\rt_{i})} \frac{\partial \by(\rt_{i})}{\partial\br(\rt_{i})} + \frac{\partial f(\by(\rt_{i}), \br(\rt_{i}))}{\partial \br(\rt_{i})} \nonumber	\\
&=& (1-\tanh^2[\cdot])  \mathbf{W}^T_{\mathrm{in}}\mathbf{W}^T_{\mathrm{out}} +  (1-\tanh^2[\cdot])  \mathbf{W}^T \nonumber \\
&=& (1-\br(\rt_{i+1})^2) \left( \mathbf{W}^T_{\mathrm{in}}\mathbf{W}^T_{\mathrm{out}} + \mathbf{W}^T \right),
\label{eq:genesnJac}
\end{eqnarray}
where from Eq.~\eqref{eq:esn_r} $\br(\rt_{i+1})^2 =  \tanh^2([\hat{\by}_{\mathrm{in}}(\rt_{i});b_\mathrm{in}]^T \, \mathbf{W}_{\mathrm{in}}+\br(\rt_{i})^T\mathbf{W})$ is the updated squared hidden state at timestep $\rt_{i+1}$. The Jacobian of the ESN is cheap to calculate as the expression $\left( \mathbf{W}^T_{\mathrm{in}}\mathbf{W}^T_{\mathrm{out}} + \mathbf{W}^T \right)$ is a constant matrix, which is fixed after the training of $\mathbf{W}_{\mathrm{out}}$. The only time-varying component is the hidden state. The Jacobian $\mbJ \in \mathbb{R}^{N_r\times N_r}$ is used for the extraction of the Lyapunov spectrum and the CLVs of a trained ESN. We time-march $D$ Lyapunov vectors $\bu_i\in\mathbb{R}^{N_r}$, and periodically perform QR decompositions, where $\bQ \in \mathbb{R}^{N_r\times D}$, and $\bR \in \mathbb{R}^{D\times D}$. The same CLV algorithm described  in Sect.~\ref{sec:CLVs} is employed to extract $D$ covariant Lyapunov vectors $\bv_i\in\mathbb{R}^{N_r}$ from a trained ESN. The pseudocode is given in algorithm \ref{alg:CLVs}.

\subsection{Validation}
\label{sec:val}

The dataset is split into three subsets, which are the training, validation, and testing subsets in a  time-ordered fashion. During training the ESN runs in open-loop, while during validation and testing, the ESN runs in closed-loop and the prediction at each step becomes the input for the next step. 
After training the ESN, its validation is necessary for the determination of the hyperparameters.  The objective is to compute the hyperparameters that minimize the logarithm of the MSE \eqref{eq:MSE}. The logarithm of the MSE is preferred because the error varies by orders of magnitude for different hyperparameters, as explained in\cite{Racca2021}. In general, instead of Eq.~\eqref{eq:MSE}, other types of error functions can be used for the hyperparamer tuning, such as the maximization of the prediction horizon \cite{Pathakchaos2017,Doan2020,Racca2021} or the minimization of the kinetic energy differences\cite{Racca2022}. Here the input scaling, $\sigma_{\mathrm{in}}$, the spectral radius, $\rho$, and the Tikhonov parameter, $\beta$, are the ESN hyperparameters that are being tuned \cite{Lukosevicius2012,Racca2022}.  In order to select the optimal hyperparameters, $\sigma_{\mathrm{in}}$ and $\rho$, we employ a Bayesian Optimization, which is a strategy for finding the extrema of objective functions that are expensive to evaluate\cite{Hoffman2011,Racca2022}. Within the optimal  $[\sigma_{\mathrm{in}},\rho]$ we perform a grid search to select $\beta$ \cite{Racca2022}. In particular, $[\sigma_{\mathrm{in}},\rho]$ are searched in the hyperparameter space $[0.1,5]\times[0.1,1]$ in logarithmic scale, while for $\beta$ we test $\{10^{-6},10^{-8}, 10^{-10},10^{-12}\}$. The Bayesian Optimization starts from a grid of $6\times6$ points in the $[\sigma_{\mathrm{in}},\rho]$ domain, and then it selects five additional points through the \texttt{gp-hedge} algorithm \cite{Hoffman2011}. We set $b_{\mathrm{in}}=1$, $d=3$ and add Gaussian noise with zero mean and standard deviation, $\sigma_n=0.0005\sigma_y$, where $\sigma_y$ is the standard deviation of the data component-wise, to the training and validation data. Adding noise to the data improves the performance of ESNs in chaotic dynamics by alleviating overfitting \cite{Vlachas2020}. A summary of the hyperparameters is shown in Table \ref{tab:ESN hyperparameters}.

\begin{table}[t]
\centering
\caption{Echo state networks' hyperparameters. Multiple values indicate that the parameter is optimized within the range.}
\renewcommand{\arraystretch}{1.2}
\begin{tabular}{l l r}	
	\hline\noalign{\smallskip}
	Parameter $\qquad$ & Name  & Value \\ 
	\noalign{\smallskip}\hline\noalign{\smallskip}
	$\rho$ & Spectral radius  & $[0.1,\,1]$ \\
	$\sigma_{\mathrm{in}}$ & Input scaling & $[0.1,\,5]$ \\
	$\beta$ & Tikhonov parameter & $\{10^{-6},10^{-8}, 10^{-10},10^{-12}\}$  \\
	$d$ & Connectivity & $3$  \\
	$b_{\mathrm{in}}$ & Input bias & $1$  \\
	$\sigma_n$ & Noise (training) $\qquad$ & $0.0005\sigma_{u}$  \\	
	\noalign{\smallskip}\hline
\end{tabular}
\label{tab:ESN hyperparameters}
\end{table}

\begin{figure}[h]
\centering
\includegraphics[width=.99\columnwidth]{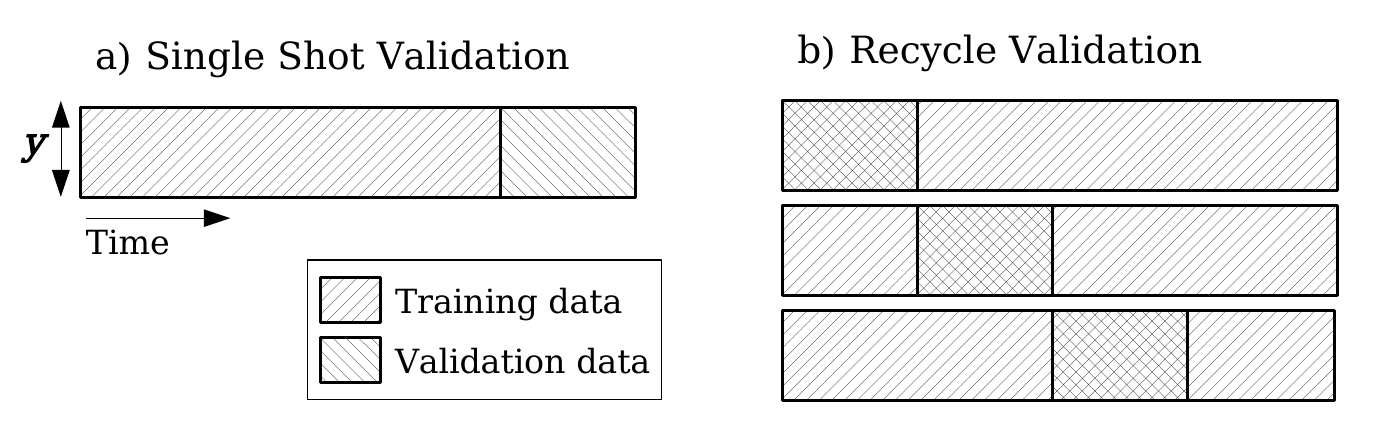}
\caption{Schematic representation of the (a) Single Shot, and (b) Recycling Validation strategies. Here, $\by$ represents the degrees of freedom of the data. Three sequential validation intervals are shown for the Recycle Validation \cite{Racca2022}.}
\label{fig:Vals}
\end{figure}

One of the most commonly used validation strategy for RNNs is the Single Shot Validation (SSV) \cite{Lukosevicius2019}, in which the data are split into a training set, followed by a single small validation set; see Fig.~\ref{fig:Vals}(a). As the ESN now runs in closed-loop, the size of the validation set is limited by the chaotic nature of the signal. In particular, at the beginning of the validation set, the input $\by(\rt_{0})$ of the ESN is initialized to the target value. However, chaos causes the predicted trajectory to quickly diverge from the target trajectory in a few Lyapunov times $\tau_\lambda$. The validation interval is therefore small and not representative of the full training set, which causes poor performance in the test set\cite{Racca2022}. An improvement to the performance with cheap computations is achieved by the the Recycle Validation (RV), which was recently proposed by\cite{Racca2022}. In the RV, the network is trained only once on the entire training dataset (in open-loop), and validation is performed on multiple intervals already used for training (but now in closed-loop); see Fig.~\ref{fig:Vals}(b). In this work we use the chaotic Recycle Validation (RVC), where the validation interval simply shifts as a small multiple of the first Lyapunov exponent, $N_{\mathrm{val}}=3\lambda_{1}$.

\section{Results}\label{sec:results}

In this section, we present the numerical results, which include a thorough comparison between the statistics produced by the autonomous temporal evolution of the ESN and the target dynamical system. The selected observables are the statistics of the degrees of freedom, the Lyapunov exponents, the angles between the CLVs or subspaces composed of CLVs, and the finite-time covariant Lyapunov exponents. We separate our analysis into two subsections, which contain two low-dimensional systems and then two higher dimensional systems.

\subsection{Low dimensional chaotic systems}\label{sec:low_dyn_syst}
As a first case, we consider two low-dimensional dynamical systems that exhibit chaotic behavior:  Lorenz 63 (L63)\cite{Lorenz63} and  R\"ossler \cite{Rossler1976} attractors. The Lorenz 63 system is a reduced-order model of atmospheric convection for a single thin layer of fluid that is heated uniformly from below and cooled from above, which is defined by  
\begin{eqnarray}
\frac{dx_1}{dt} &=& \sigma(x_2-x_1) \nonumber\\
\frac{dx_2}{dt} &=& x_1(\rho-x_3) -x_2 \\
\frac{dx_3}{dt} &=& x_1 x_2 - \beta x_3. \nonumber\label{eq:l63}
\end{eqnarray}
We chose the parameters $[\sigma, \beta, \rho] = [10, 8/3, 28]$ to ensure a chaotic behavior. The R\"ossler attractor, 
which models equilibrium in chemical reactions, is governed by 
\begin{eqnarray}
\frac{dx_1}{dt} &=& -(x_2 + x_3) \nonumber\\
\frac{dx_2}{dt} &=& x_1 +ax_2 \\
\frac{dx_3}{dt} &=& b +x_3(x_1-c),  \nonumber\label{eq:rossler}
\end{eqnarray}
We choose the parameters $[a, b, c] = [0.1, 0.1, 14]$ to ensure a chaotic behavior.

To generate the target set we evolve the dynamical systems forward in time with a 4$^{\text{th}}$ order Runge-Kutta (RK4) integrator and a timestep $dt= 0.005$ for both L63 and R\"ossler, which is sufficiently small for a good temporal resolution. (We tested slightly larger/smaller timesteps with no significant differences. Results not shown.) We perform a QR decomposition every $m=1$ timesteps for L63 and every $m=5$ timesteps for R\"ossler. For all systems, we generate a training set of size $1000\tau_\lambda$, and a test set of size $4000\tau_\lambda$, for the CLV statistics to converge, where $\tau_\lambda=1/\lambda_1$ is the Lyapunov time, which is the inverse of the maximal Lyapunov exponent $\lambda_1$.

\begin{figure}[t]
\centering
\includegraphics[width=.99\columnwidth]{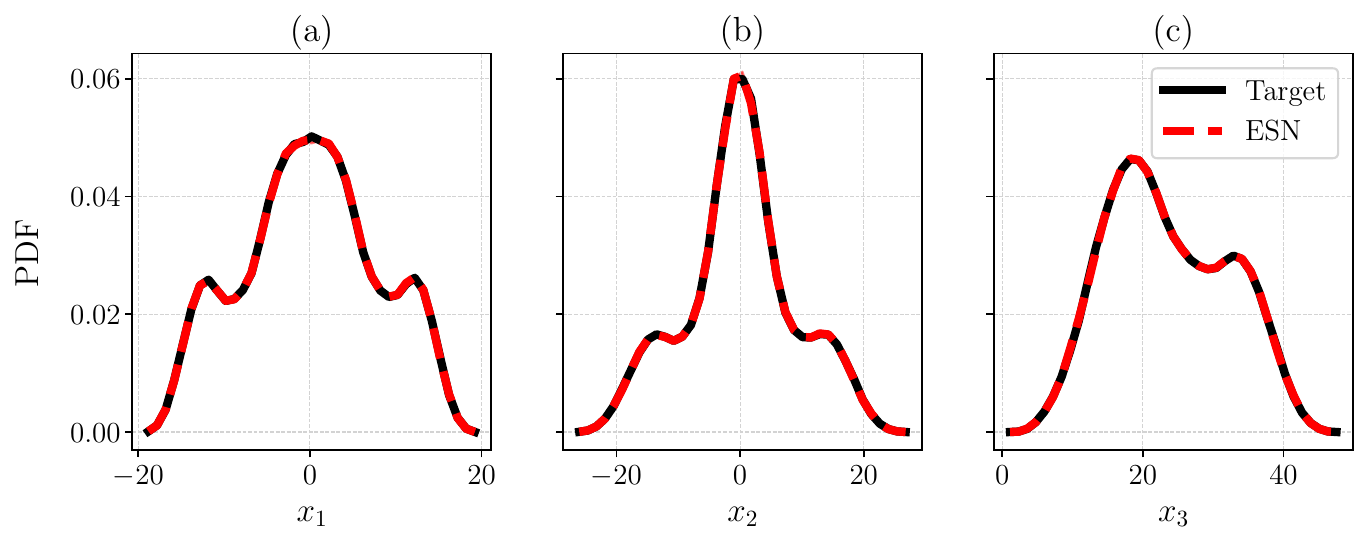}
\includegraphics[width=.99\columnwidth]{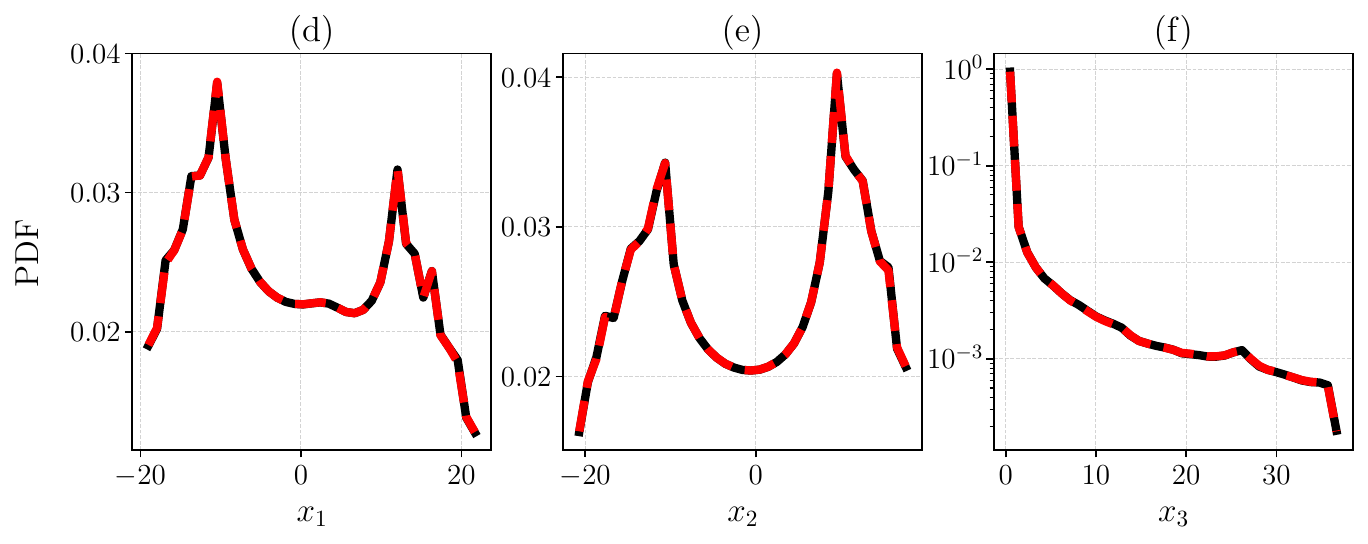}
\caption{Comparison of the Target (straight black line) and ESN (red dashed line) Probability Density Functions (PDF) of the three degrees of freedom, $x_1$, $x_2$, and $x_3$ of the Lorenz 63 system \eqref{eq:l63} (a-c) and the R\"ossler system \eqref{eq:rossler} (d-f).}
\label{fig:L63_Ross_variables_pdf}
\end{figure}

\begin{figure}[h]
\centering
\includegraphics[width=.99\columnwidth]{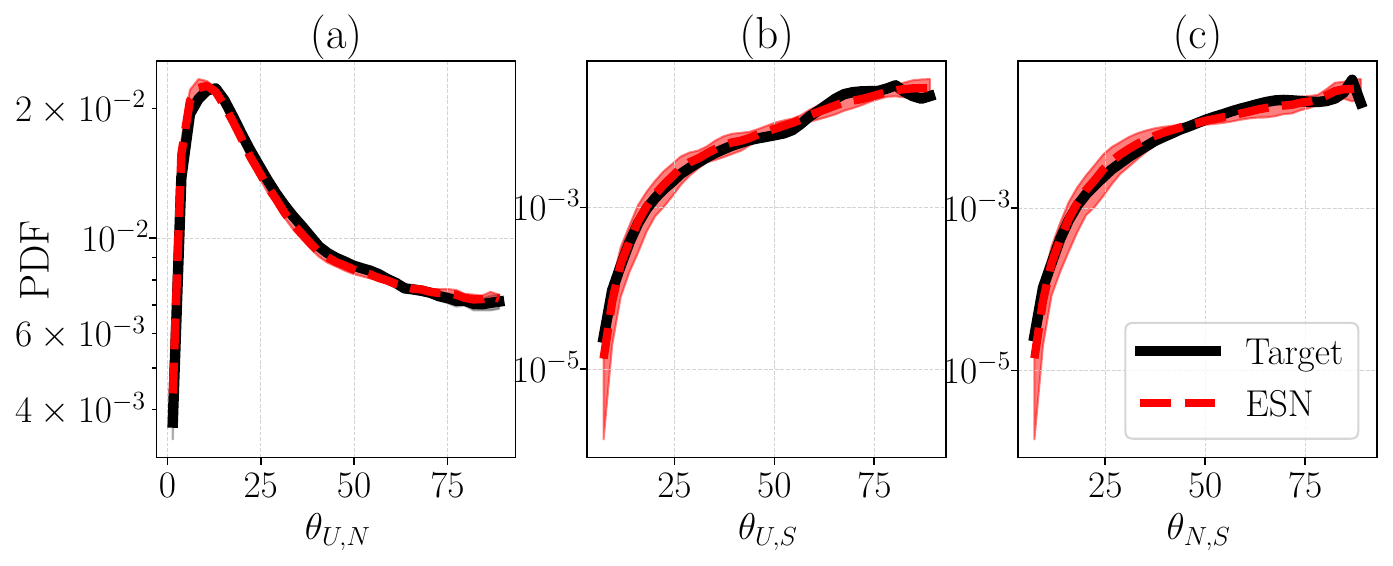}
\includegraphics[width=.99\columnwidth]{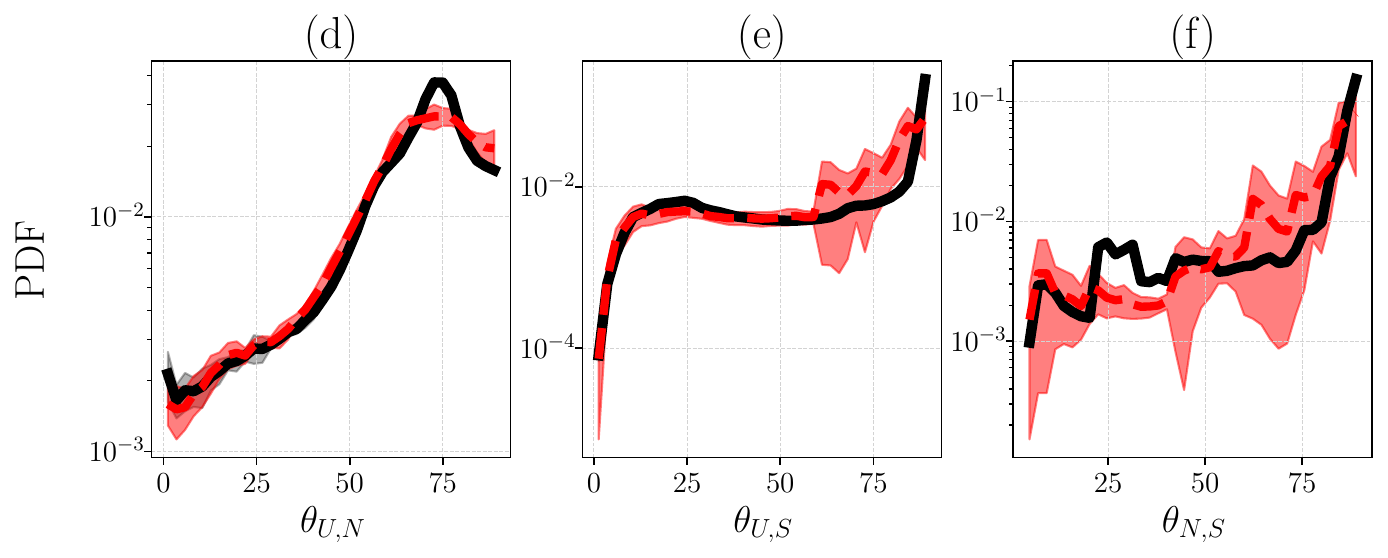}
\caption{Comparison of the Target (straight black line) and ESN (red dashed line) Probability Density Functions (PDF) of the three principal angles between the Covariant Lyapunov Vectors, where $U$ refers to unstable, $N$ to Neutral and $S$ to stable CLVs. The top row (a-c) is for Lorenz 63 \eqref{eq:l63}  and the bottom row (d-f) for R\"ossler \eqref{eq:rossler}. All $y$-axes are in logarithmic scale and the $x$-axis is in degrees. The shaded region indicates the error bars derived by the standard deviation.}
\label{fig:L63_Ross_thetas_pdf}
\end{figure}

\begin{figure}[h]
\centering
\includegraphics[width=.99\columnwidth]{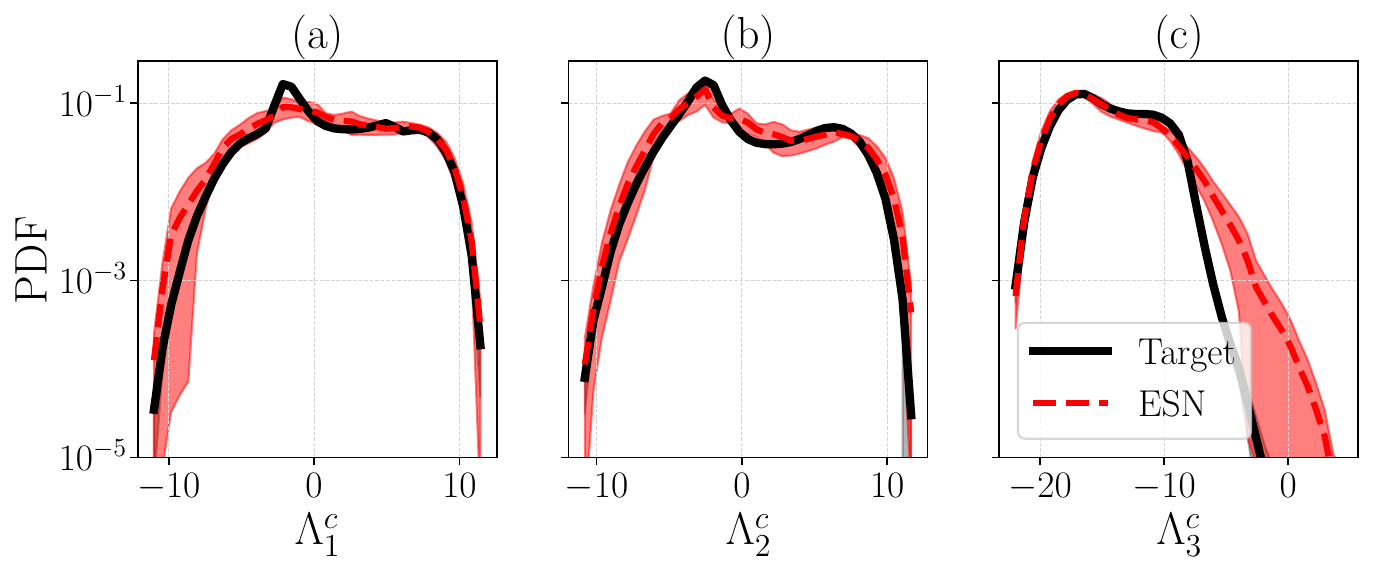}
\includegraphics[width=.99\columnwidth]{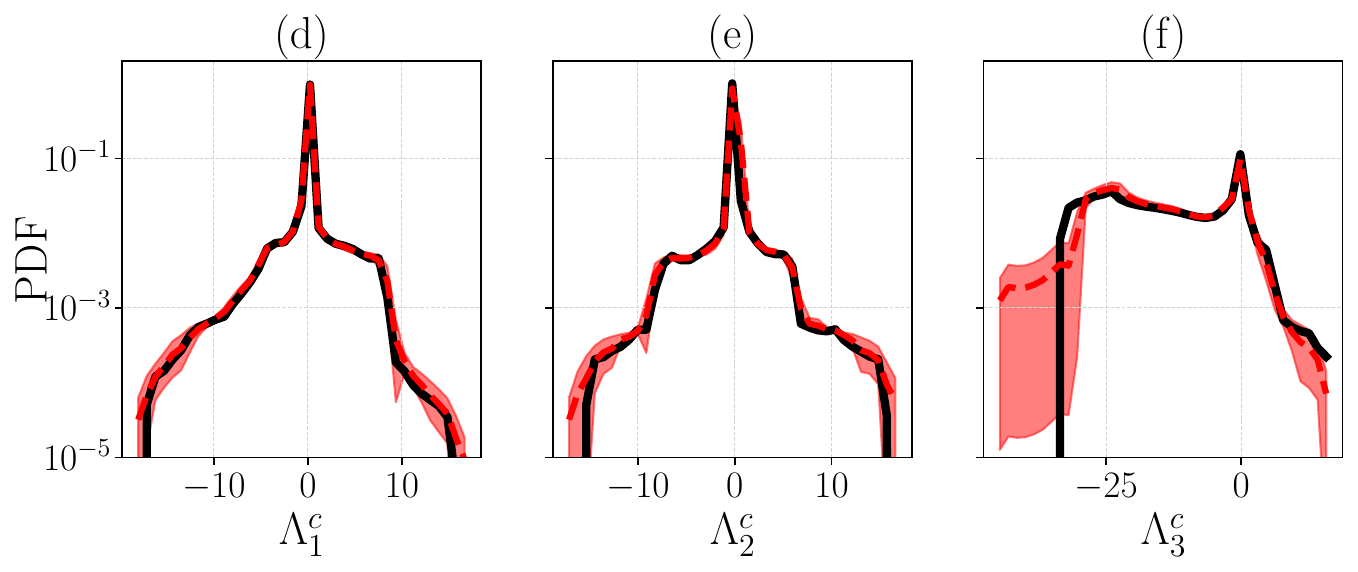}
\caption{Comparison of the Target (straight black line) and ESN (red dashed line) Probability Density Functions (PDF) of the three finite-time Covariant Lyapunov Exponents. The top row (a-c) is for Lorenz 63 \eqref{eq:l63}  and the bottom row (d-f) for R\"ossler \eqref{eq:rossler}. All $y$-axes are in logarithmic scale.}
\label{fig:L63_Ross_icle_pdf}
\end{figure}

First, we test whether the ESN correctly learns the chaotic attractor from a statistical point of view, i.e., whether the ESN correctly learns the long-term statistics of the degrees of freedom when it evolves in the closed-loop (autonomous) mode. By estimating the probability density function (PDF) of the degrees of freedom of the ESNs, as a normalized histogram, and comparing it with the corresponding PDF of the target set, we extract information on the invariant measure of the considered chaotic system. This is shown in Fig.~\ref{fig:L63_Ross_variables_pdf} for L63 and R\"ossler attractors, in which the black lines show the target statistics and the red dashed lines show the ESN statistics.
In Figs.~\ref{fig:L63_Ross_variables_pdf},~\ref{fig:L63_Ross_thetas_pdf}, \ref{fig:L63_Ross_icle_pdf}, and Table ~\ref{tab:LEs_low_dim} we have used $N_{\mathrm{ESN}}$ ESNs trained on $N_{\mathrm{ESN}}$ independent target systems, starting from different initial conditions, and averaged among the estimated observables, where $N_{\mathrm{ESN}}=6$, and $N_{\mathrm{ESN}}=8$ for  R\"ossler and L63, respectively. We  perform the ensemble calculation to quantify the uncertainty of the predictions and the robustness of the ESN for different initializations.

Second, we test whether the ESN correctly learns the Lyapunov spectrum. Table~\ref{tab:LEs_low_dim} shows the ESN predictions on the Lyapunov exponents for the L63 and R\"ossler attractors, which are compared with the target exponents. The leading exponent is accurately predicted with a 0.2\% error in the L63, and 1.5\% error in the R\"ossler system. In chaotic systems, there  exists a neutral Lyapunov exponent, which is associated with the direction of $\frac{d\bx}{dt}$. In these cases the neutral Lyapunov exponents are $\lambda_2 = 0$ for both systems, which are correctly inferred by the ESN within a $\mathcal{O}(10^{-5})$ error, or less. For the smallest, and negative exponent, which is generally harder to extract because it is highly damped, the relative error is about 0.6\% for L63 and 2.1\% for R\"ossler. Therefore, the ESNs can accurately capture the tangent dynamics of a low dimensional chaotic attractor.

\begin{table}[]
\caption{Estimates of Lyapunov exponents $\lambda_{i}$ for the two low dimensional systems, the Lorenz 63 and R\"ossler attractors. Comparison between the target and echo state network.}
\renewcommand{\arraystretch}{1.2}
\centering
\begin{tabular}{ccc|cc}\hline
	& \multicolumn{2}{c|}{Lorenz 63} & \multicolumn{2}{c}{R\"ossler} \\ \cline{1-5} 
	\multicolumn{1}{c|}{$\lambda_{i}$}	& target        &  ESN     & target     &  ESN    \\ \hline
	\multicolumn{1}{c|}{1} & 0.9050 & 0.9067 & 0.071 & 0.070  \\
	\multicolumn{1}{c|}{2} & 9$\times10^{-5}$ &  -8.$\times10^{-5}$  &  $2\times10^{-6}$ & $1\times10^{-6}$           \\
	\multicolumn{1}{c|}{3} & -14.572 & -14.664 & -13.88 & -14.17         \\ \hline
\end{tabular}
\label{tab:LEs_low_dim}
\end{table}

Third, we investigate the angles between the CLVs. We assess whether the ESNs learn the long-term statistics of these quantities, but also whether, they correctly infer the distribution and fluctuations of those observables in the phase space. In other words, whether the ESNs learn the geometrical structure of the attractor and its tangent space. 

\begin{figure}[t]
\centering
\includegraphics[width=.55\textwidth]{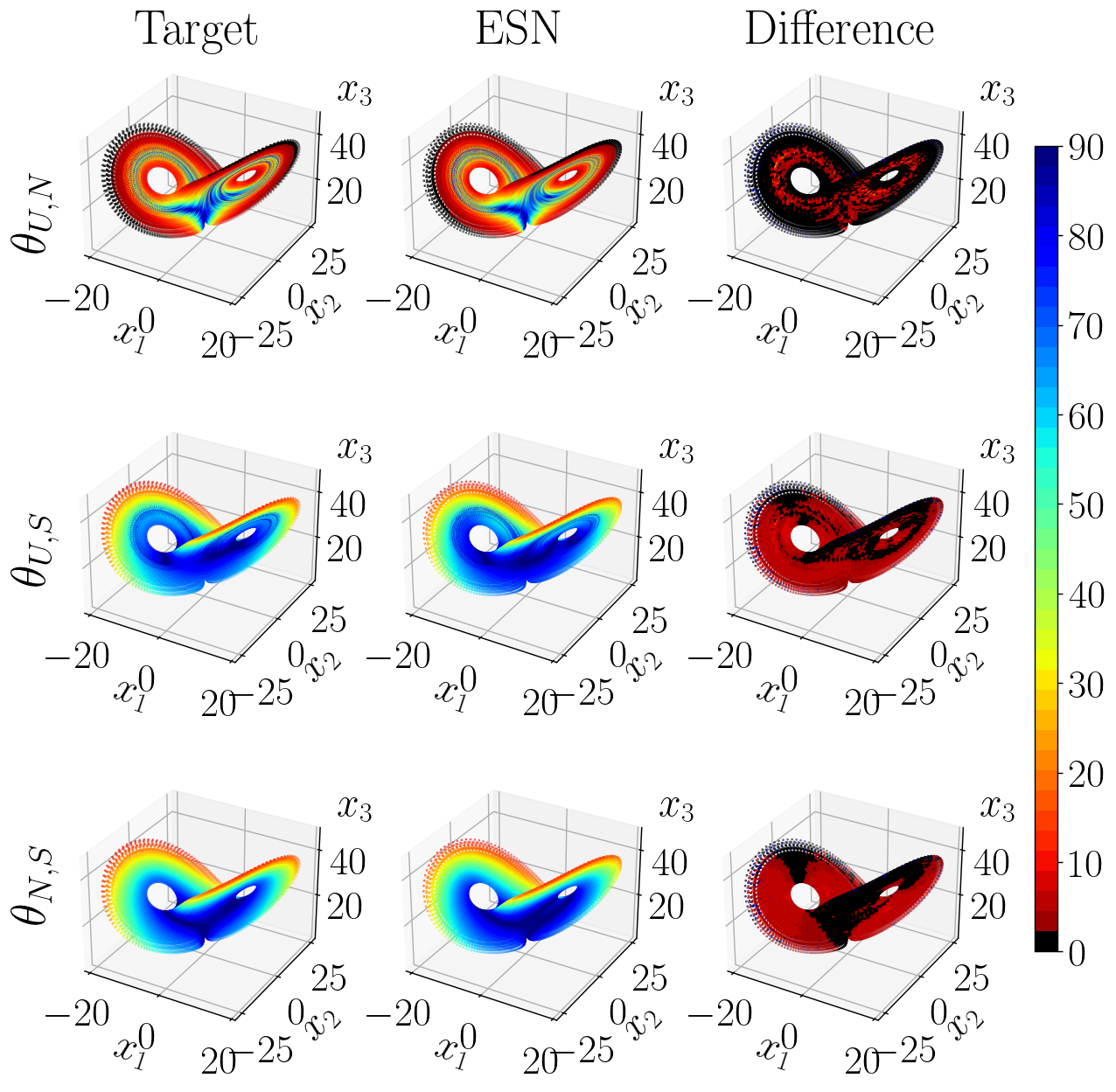}
\caption{Comparison of Target (left column), ESN (middle column), and their statistical mean absolute difference (right column), for a $300\tau_\lambda$ trajectory of the Lorenz 63 system \eqref{eq:l63} in the test set, coloured by the CLV principal angles (in $\deg$). First row: $\theta_{U,N}$, second row: $\theta_{U,S}$, and third row: $\theta_{N,S}$.}
\label{fig:L63_theta_att}
\end{figure}

In Fig.~\ref{fig:L63_theta_att}, we present an analysis of the distribution of principal angles between the CLVs,
\begin{equation}
\theta_{a,b} = \frac{180^{\circ}}{\pi} \cos^{-1}(|\bv_a\cdot \bv_b|),
\label{eq:clvangles}
\end{equation} 
$\theta_{a,b} \in [0^{\circ},90^{\circ}]$, on the topology of the L63 attractor. The attractor is well reproduced by a selected ESN (middle column), compared to the target (left column). The size of both trajectories is $300\tau_\lambda$.  In this case, there are three principal angles between the CLVs; $\theta_{U,N}$ is the angle between the unstable and neutral CLV; $\theta_{U,S}$ is the angle between the unstable and stable CLV; $\theta_{N,S}$ is the angle between the neutral and  stable CLV.  The colouring of the attractor is associated with the measured $\theta_{a,b}$. The black and dark red colours identify small angles, i.e., regions of the attractor where near-tangencies between the CLVs occur. Possible tangencies between CLVs or invariant manifolds composed of CLVs (as will be discussed later for higher dimensional chaotic systems) are of significant importance, as they signify that the attractor is non-hyperbolic\cite{Ginelli2013} (see Sect.~\ref{sec:CLVs}).  The right column is the  mean absolute difference between the target and the ESN. 
The $x,y,z$ domain is discretized with 50 bins in each direction; then the mean $\theta_{a,b}$ is calculated from each of the three-dimensional bins for the $300\tau_\lambda$ long trajectory. Finally, the absolute  difference between ESN and target is calculated for each bin. The plots follow the same color scheme as the colorbar, with black and dark red colors indicating $<2^\circ$ differences with a maximum of $\sim10^\circ$. Figure~\ref{fig:L63_theta_att} shows that the ESN is able to accurately learn the dynamics of the tangent linear space of the attractor.  

In Fig.~\ref{fig:L63_Ross_thetas_pdf}, we show the PDF of the  principal angles between the three CLVs, for which there is agreement between target and ESN results in all cases for both L63 and R\"ossler, even for smaller angles. The non-zero count of events close to $\theta\to0$ indicates that the two considered systems are non-hyperbolic, which is consistent with the literature\cite{kuptsov2018}.

Fourth, we analyse the distribution on the attractor, as well as the statistics, of the Finite-time Covariant Lyapunov Exponents, for a time-lapse of $\Delta t = m \,dt$ timestep , and assess the accuracy of the trained ESNs. For the considered low-dimensional systems there are three FTCLEs with each showing the finite-time growth rate of the corresponding Covariant Lyapunov Vectors.

\begin{figure}[t]
\centering
\includegraphics[width=.56\textwidth]{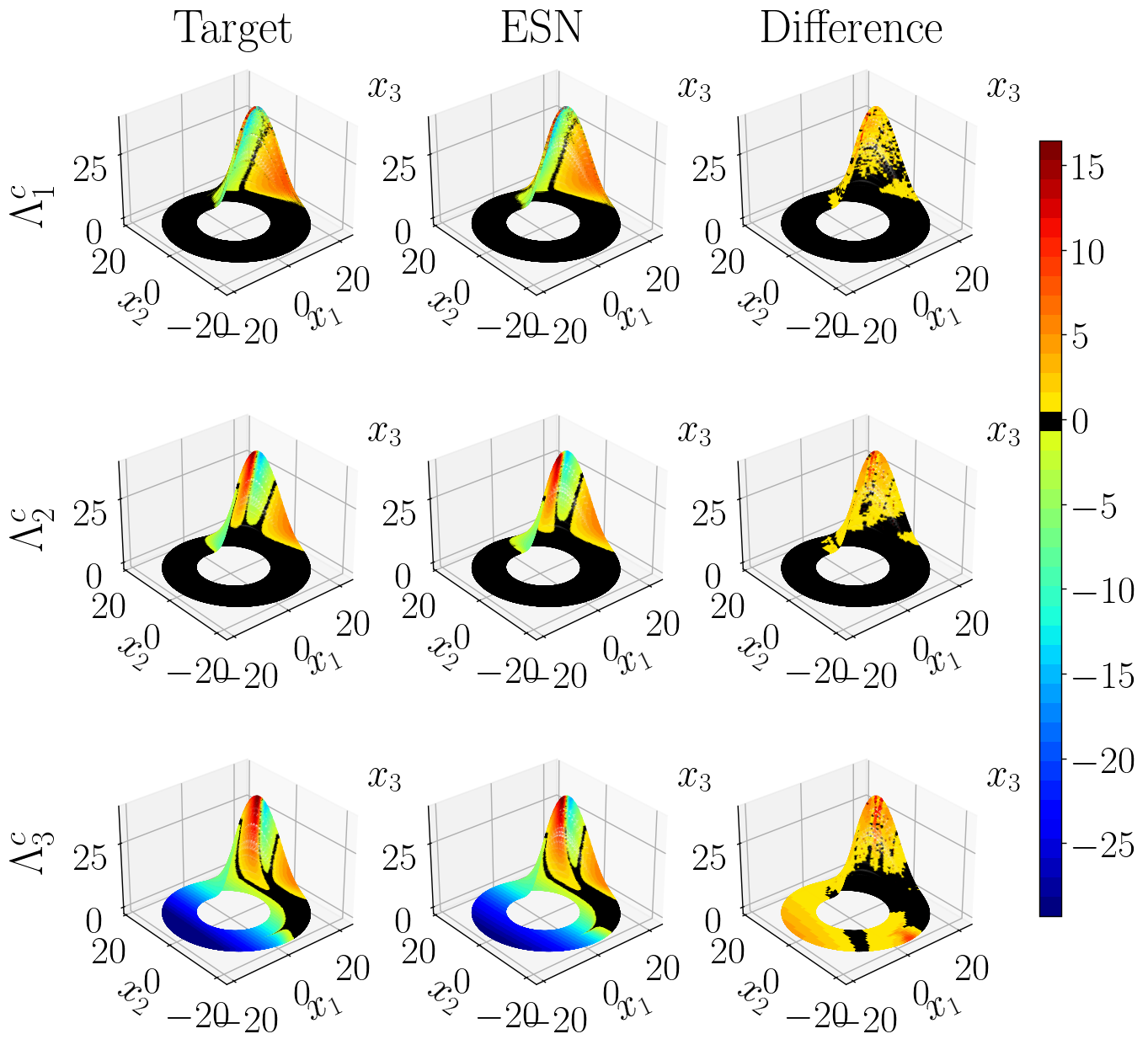}
\caption{Comparison of Target (left column), ESN (middle column), and their statistical mean absolute difference (right column) for a $300\tau_\lambda$ trajectory of the R\"ossler system \eqref{eq:rossler} in the test set, coloured by the three FTCLEs. First row: FTCLE 1, second row: FTCLE 2, and third row: FTCLE 3.}
\label{fig:Rossler_icle_att}
\end{figure}

In Fig.~\ref{fig:Rossler_icle_att} we visualize the distribution of the single timestep FTCLEs, in the case of the R\"ossler attractor, which is well reproduced by a selected ESN (middle column), compared to the target (left column). The size of both trajectories is $300\tau_\lambda$. FTCLE 1 is the finite-time exponent for the unstable CLV, FTCLE 2 is for the neutral CLV, and FTCLE 3 is for the stable CLV. The colouring is associated with the values of the FTCLEs. Large positive FTCLEs correspond to high finite-time growth rates and, thus, reduced predictability. The distribution of the leading FTCLE on the attractor is similar between the target and ESN. 
The second FTCLE and third FTCLE, which correspond to the neutral CLV and stable CLVs, accordingly, also show good agreement between the two. The  mean difference between the target and the ESN on the attractor is plotted in the right column, in which black identifies $\Lambda_i^c\approx0$. The right column shows that most of the small differences between the ESN and the target are  located in the region of  large variation of $z$.

Finally, Fig.~\ref{fig:L63_Ross_icle_pdf} shows the PDF of the three FTCLEs. There is agreement between the ESN-inferred quantities and the target in all cases, in particular in the R\"ossler attractor for the most-probable statistics.
The small  deviation in Fig.~\ref{fig:L63_Ross_icle_pdf}(a) for L63 corresponds to the statistics around the peak of the first FTCLE, $\Lambda_1^c$, but the tails of the distributions are well reproduced.
The mean of the $\Lambda_i^c$ distributions coincides with the LEs $\lambda_{i}$, which holds true for all our results. A behavior as in Fig.~\ref{fig:L63_Ross_icle_pdf}(a) implies that in this case the finite-time values $\Lambda_1^c$ are less peaked around the mean value, even though their long-time average coincides with the Lyapunov exponent $\lambda_{1}$. Nevertheless, in Figs.~\ref{fig:L63_Ross_icle_pdf}(d-f) for R\"ossler the statistics around the peak (and beyond) are  well captured. 

We refer the interested reader to our supplementary material where the corresponding results of Figs.~\ref{fig:L63_theta_att},\ref{fig:Rossler_icle_att} for both attractors are shown. Also, the statistics of FTLEs, as well as their distribution on the chaotic attractors are presented in the supplementary material.

\begin{figure*}[t]
\centering
\includegraphics[width=.9\textwidth]{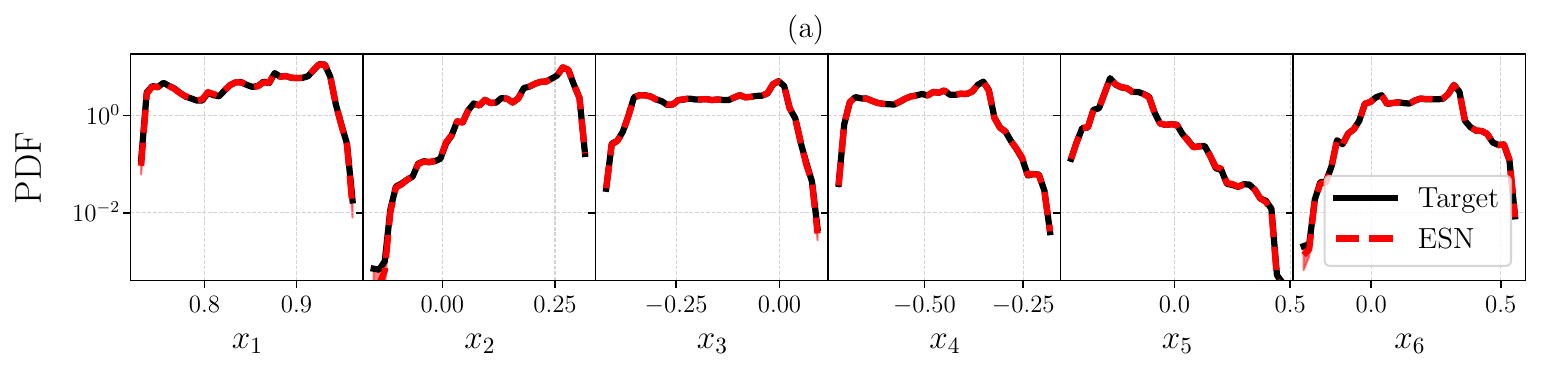}
\includegraphics[width=.9\textwidth]{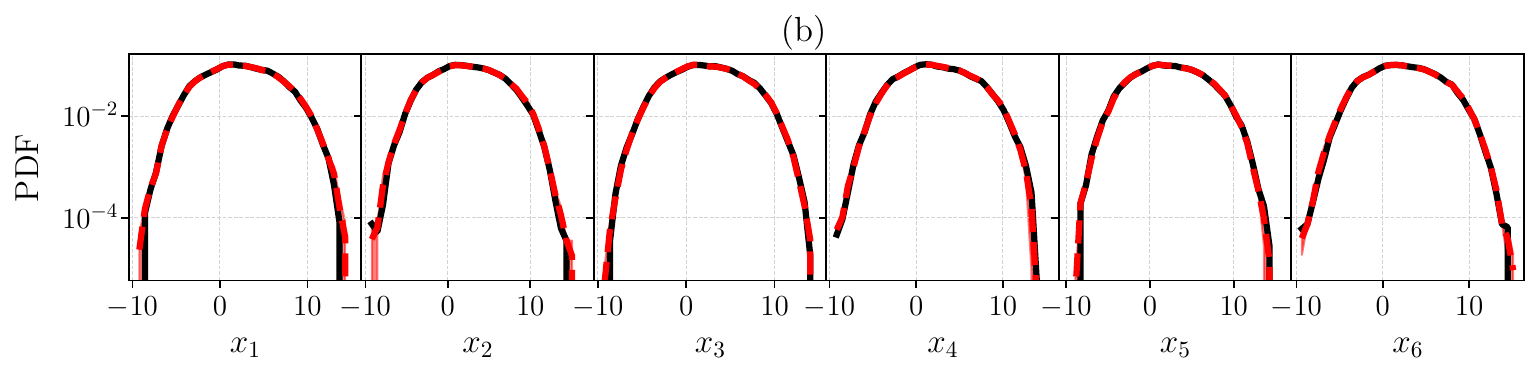}
\caption{Comparison of the Target (straight black line) and ESN (red dashed line) PDF of the first six degrees of freedom, $x_1$, \dots, $x_6$ of the (a) Charney-DeVore \eqref{eq:cdv}  and (b) for Lorenz 96 \eqref{eq:l96} for $D=20$.}
\label{fig:L96_pdf_dof}
\end{figure*}

\subsection{Higher dimensional chaotic systems}\label{sec:high_dyn_syst}

We follow the same analysis and approach as in Sect.~\ref{sec:low_dyn_syst} for two  higher dimensional chaotic systems, both of which are related to atmospheric physics and meteorology. The first is a reduced-order model of atmospheric blocking events by Charney and DeVore\cite{CdV1979} (CdV), which is a six-dimensional truncation of the equations for barotropic flow with orography. 
We employ the formulation of \cite{DeSwart1989,Crommelin2004}, which is forced by a zonal flow profile that can be barotropically unstable. The governing equations are 
\begin{align}
\frac{dx_1}{dt} &= \gamma_1^*x_3 - C(x_1 - x_1^*), \nonumber\\ \nonumber
\frac{dx_2}{dt} &= -(\alpha_1 x_1 - \beta_1)x_3 - C x_2 - \delta_1x_4x_6,\\ 
\frac{dx_3}{dt} &=  (\alpha_1 x_1 - \beta_1)x_2 - \gamma_1 x_1 - C x_3 + \delta_1 x_4 x_5,\\ \nonumber
\frac{dx_4}{dt} &=  \gamma_2^*x_6 - C(c_4-x_4^*) + \epsilon( x_2 x_6 - x_3x_5),\\ \nonumber
\frac{dx_5}{dt} &= -(\alpha_2 x_1 - \beta_2)x_6 - C x_5 - \delta_2 x_4 x_3,\\ \nonumber
\frac{dx_6}{dt} &=  (\alpha_2 x_1 - \beta_2)x_2 - \gamma_2 x_4 - C x_6 + \delta_2 x_4 x_2, 
\label{eq:cdv}
\end{align}
where the model coefficients are  
\begin{align}
\alpha_m &= \frac{8\sqrt{2}}{\pi} \frac{m^2}{4m^2-1}\frac{b^2+m^2-1}{b^2+m^2}, \quad \quad \beta_m = \frac{\beta b^2}{b^2+m^2},\nonumber\\ 
\delta_m &= \frac{64\sqrt{2}}{15\pi} \frac{b^2-m^2+1}{b^2+m^2}, \quad \quad  \quad\gamma^*_m = \gamma \frac{4m}{4m^2-1}\frac{\sqrt{2}b}{\pi},\\ \nonumber
\epsilon &= \frac{16\sqrt{2}}{5\pi}, \quad \quad \quad \quad \quad \quad \gamma_m = \gamma \frac{4m^3}{4m^2-1}\frac{\sqrt{2}b}{\pi(b^2+m^2)}. 
\label{eq:cdv_params}
\end{align}
Eq.~\eqref{eq:cdv} is integrated with RK4 and $dt = 0.1$. The constants are set to $(x_1^*, x_4^*,C,\beta,\gamma,b) = (0.95, -0.76095, 0.1, 1.25, 0.2, 0.5)$, for which the CdV model generates regime transitions\cite{DeSwart1989,Crommelin2004}.  In particular, the CdV model allows for two metastable states, the so-called ``zonal'' state, which represents the approximately zonally symmetric jet stream in the mid-latitude atmosphere, and the ``blocked'' state, which  refers to a diverse class of weather patterns that are a persistent deviation from the zonal state. Blocking events are known to be associated with regional extreme weather, from heatwaves in summer to cold spells in winter\cite{Woollings2018}. The dynamical properties of CLVs in connection to blocking events were recently investigated for a series of more complex atmospheric models than CdV \cite{Vannitsem2016,Schubert2015,Schubert2016}, which demonstrated that CLVs are good candidates for blockings precursors, as wells as a good basis for model reduction. In previous work\cite{Doan2020}, the CdV system was used as a training model for the ESN, with the purpose of studying short-term accurate prediction of chaos, and quantifying the benefit of  Physics informed Echo State Networks\cite{Doan2020}.

The second higher-dimensional system that we consider is the Lorenz 96 (L96) model\cite{Lorenz96}, which is a system of coupled ordinary differential equations that describes the large-scale behavior of the mid-latitude atmosphere, and the transfer
of a scalar atmospheric quantity. Three characteristic processes of atmospheric systems (advection, dissipation, and external forcing) are included in the model, whose equations are 
\begin{equation}
\frac{d\bx_i}{dt} = (\bx_{i+1} - \bx_{i-2})\bx_{i-1} - \bx_i + F,
\label{eq:l96}
\end{equation}
where $\bx = [ x_1, x_2, \dots,x_D ] \in \mathbb{R}^D$.  We set periodic boundary conditions, i.e.~$x_{1}=x_{D+1}$. In our analysis we chose $D=20$ degrees of freedom. The external forcing is set to $F=8$, which ensures a chaotic evolution\cite{Vlachas2020}. We integrate the system with RK4 and $dt = 0.01$. We perform a QR decomposition every $m=5$ timesteps for CdV and every $m=10$ timesteps for L96. Similar to the previous section, we generate a training set of size $1000\tau_\lambda$ and a test set of size $4000\tau_\lambda$.

First, Fig.~\ref{fig:L96_pdf_dof} shows the PDF of the six degrees of freedom of CdV, and the first six from L96 (the PDFs of the rest 13 dofs have similar shape and agreement between ESN and target). We use a semilogarithmic scale to emphasize that the agreement between target (black line) and ESN (red dashed line) is accurate for the tails of the distributions, which effectively correspond to the edges of each attractor. As in Sect.~\ref{sec:low_dyn_syst} in order to evaluate uncertainty and robustness, we start with $N_{\mathrm{ESN}}=10$ trained networks, but during post-processing we discard any network that shows spurious temporal evolution, and perform a further averaging of the PDFs of each network's observable. Therefore, the PDFs of Fig.~\ref{fig:L96_pdf_dof} are the outcome of averaging $N_{\mathrm{ESN}}=5$ and $N_{\mathrm{ESN}}=9$ PDFs with the same binning, for CdV and L96, respectively.

\begin{figure}[h]
\centering
\includegraphics[width=.9\columnwidth]{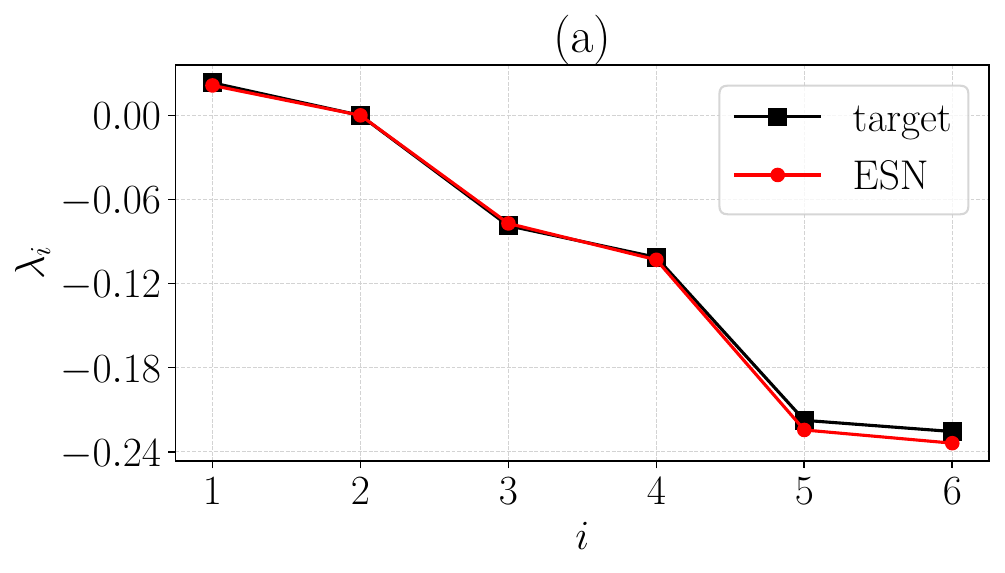}
\includegraphics[width=.9\columnwidth]{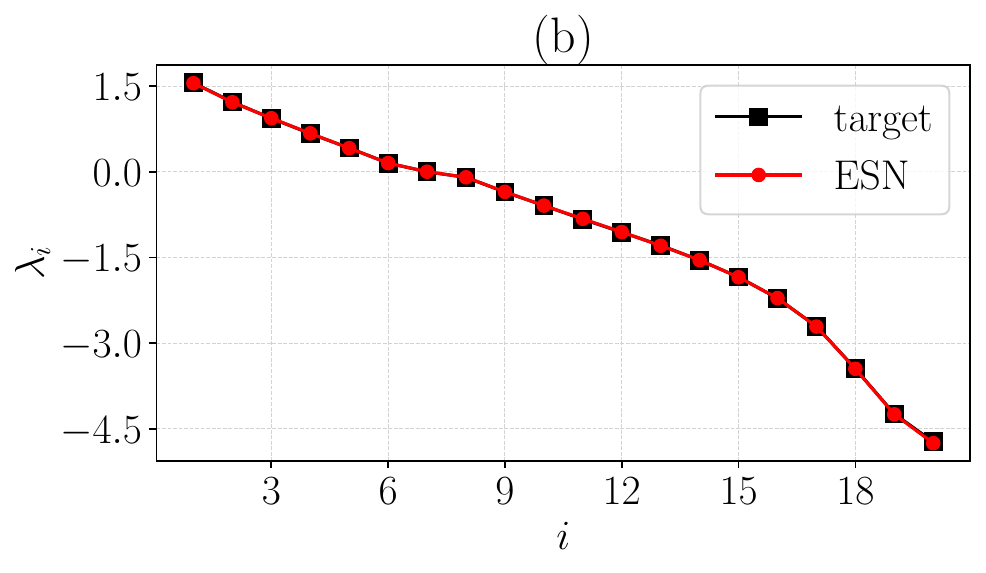}
\caption{Comparison of the Target (black squares) and ESN (red circles) Lyapunov spectrum for (a) Charney-DeVore \eqref{eq:cdv},  and (b) Lorenz 96 \eqref{eq:l96} at $D=20$.}
\label{fig:L96_cdv_LEs}
\end{figure}

Second, Fig.~\ref{fig:L96_cdv_LEs} shows the Lyapunov Exponents spectrum of (a) CdV and (b) L96 for $D=20$, and compares the target (black squares) with the ESN prediction (red circles). The CdV model has a single positive Lyapunov exponent, with the average value of 5 ESNs resulting in $\lambda_1 = 0.0214$, and for the 5 independent target sets, $\lambda^{\text{targ}}_1 = 0.0232$ with an $8\%$ absolute error. The second Lyapunov exponent is zero (to numerical error), and corresponds to the neutral direction, with $\lambda_2 = -3\times10^{-5}$ for ESN, and $\lambda^{\text{targ}}_2 = -7\times10^{-6}$ for the target. The low order of magnitude achieved by the ESN assures its ability to capture the neutral exponent. Finally, the four remaining negative exponents are well learned by the ESN, i.e.,  $\lambda_{3-6} =[-0.077$, $-0.103$, $-0.224$, $-0.234]$ and $\lambda^{\text{targ}}_{3-6} =[-0.079$, $-0.101$, $-0.218$, $-0.226]$ for the target. Overall, excluding $\lambda_2$, the mean absolute error of the CdV Lyapunov spectrum here is 3.7\%, which is negligibly small. 

With respect to the L96 Lyapunov spectra in Fig.~\ref{fig:L96_cdv_LEs}(b), the agreement between target and ESN across all 20 exponents is good. In particular, there are 6 positive, 1 zero and 13 negative exponents. The maximal exponent predicted from the ensemble of $N_{\mathrm{ESN}}=9$ ESNs is equal to $\lambda_1 = 1.551$, and for the 9 independent target sets, $\lambda^{\text{targ}}_1 = 1.557$, meaning a $0.4\%$ absolute error. The rest of the positive exponents are  well captured by the ESN, with $\lambda_{2-6} =[1.221$, $0.936$, $0.668$, $0.416$, $0.151]$ and $\lambda^{\text{targ}}_{2-6} =[1.217$, $0.937$, $0.673$, $0.413$, $0.152]$ for the target. The zero exponent is sufficiently small with $\lambda_7 = -10^{-4}$ for ESN, and $\lambda^{\text{targ}}_7 = 4\times10^{-4}$ for target. Albeit more difficult to predict because of large numerical dissipation, the negative Lyapunov exponents are accurately learned by the ESN, with the smallest ones reading $\lambda_{15-20} =[-1.84$, $-2.22$, $-2.71$,  $-3.45$, $-4.24$, $-4.73]$ and accordingly $\lambda^{\text{targ}}_{15-20} =[-1.85$, $-2.21$, $-2.71$,  $-3.45$, $-4.25$, $-4.75]$ for the target. Those directions in tangent space decay exponentially fast and the accuracy that the ESN achieves is consistent. For L96 the mean absolute error of the Lyapunov spectrum is approximately 0.5\%.

\begin{figure}[t]
\centering
\includegraphics[width=1.\columnwidth]{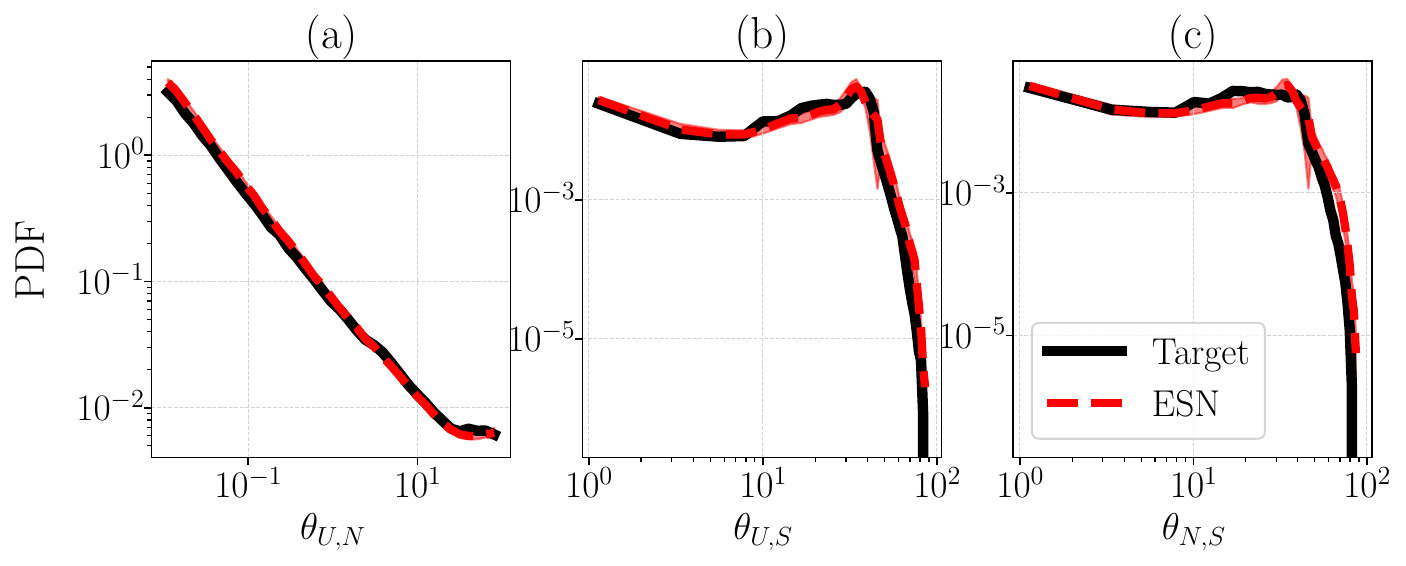}
\includegraphics[width=1.\columnwidth]{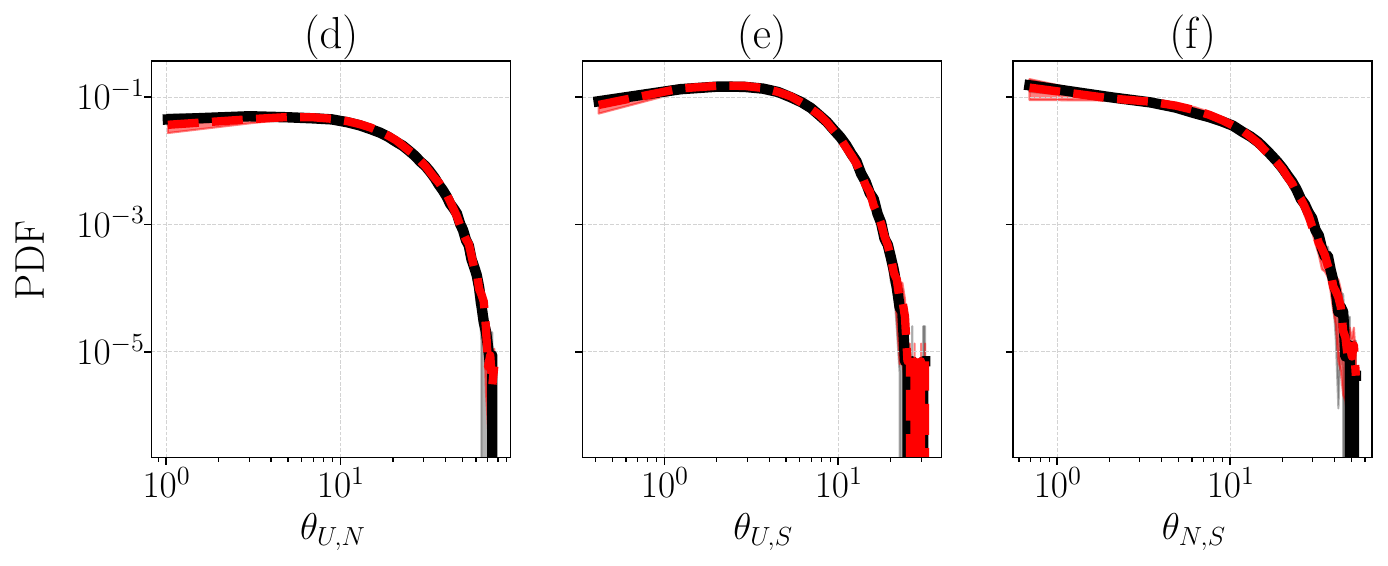}
\caption{Comparison of the Target (straight black line) and ESN (red dashed line) PDF of the three minimum principal angles between the three subspaces composed by the CLVs, where $U$ refers to unstable, $N$ to neutral and $S$ to stable CLVs. The top row (a-c) is for Charney-DeVore \eqref{eq:cdv} and the bottom row (d-f) for Lorenz 96 \eqref{eq:l96} at $D=20$. Both $x$ and $y$ axes are in logarithmic scale and the $x$-axis is in degrees. Only in (a) a logarithmic binning was used being denser close to $\theta_{U,N}\to0$, while PDFs in (b-f) are linearly binned in $x$-axis. }
\label{fig:L96_cdv_theta}
\end{figure}

\begin{figure*}[t]
\centering
\includegraphics[width=.9\textwidth]{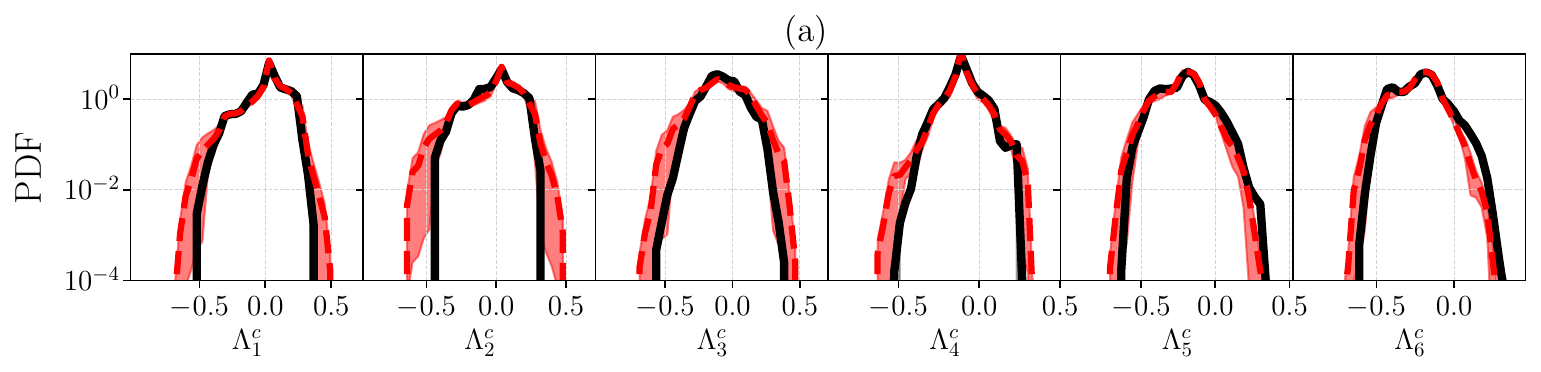}
\includegraphics[width=.9\textwidth]{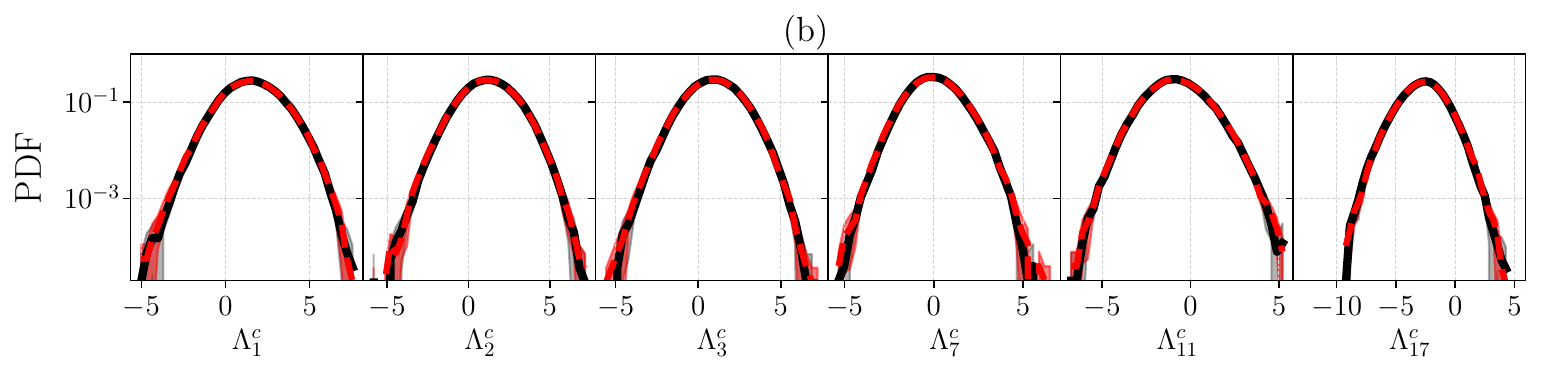}
\caption{Comparison of the Target (straight black line) and ESN (red dashed line) PDF of six Finite time Covariant Lyapunov Exponents ($\Lambda^c_i$) for (a) Charney-DeVore \eqref{eq:cdv} for which $\lambda_1>0$,  $\lambda_2=0$  and the rest are $\lambda_i<0$, and (b) Lorenz 96 \eqref{eq:l96} at $D=20$, where for $i=1,2,3$ $\lambda_i>0$, for $i=7$  $\lambda_i=0$, and for $i=11,17$ $\lambda_i<0$. All $y$-axes are in logarithmic scale.}
\label{fig:L96_pdf_icle}
\end{figure*}
To further elaborate, the L96 is known to be an extensive system \cite{Ruelle1982large,Karimi2010}, which means that quantities such as the surface width, the entropy and the attractor dimension scale linearly with its dimensionality $D$. For the Lyapunov spectrum this means that the proportion of positive to negative exponents is roughly the same ($\approx 1/2$) as $D$ changes. 
For this reason, our chosen $D=20$ is sufficient for our purposes. 

Third, we investigate the statistics of the principal angles, $\theta\in[0^{\circ},90^{\circ}]$, between the three subspaces that partition the invariant manifolds, which are the unstable $\bE^U_\bx$, neutral $\bE^N_\bx$ and stable $\bE^S_\bx$, spanned by the corresponding CLVs. 
The extraction of the principal angles between two linear subspaces requires a singular value decomposition of their matrix product $\mGamma^{a,b} = \bE^a_\bx \bE^b_\bx$ (assuming the CLVs are ordered as stacked columns, according to their Lyapunov exponent order), because all paired products between the CLVs spanning the subspaces do not provide all the angles\cite{Kuptsov2009,Ginelli2013}. The angles are given by
\begin{equation}
\theta_{a,b} = \frac{180^{\circ}}{\pi}  \cos^{-1}\left({\text{svd}[\mGamma^{a,b}]}\right),
\label{eq:svdangles}
\end{equation}
and we analyse the smallest singular value. Here, we use the implemented routine \texttt{scipy.linalg.subspace\_angles} of the scipy package\cite{SciPy2020} in python, and analyse the minimum angle in order to track homoclinic tangencies between the subspaces. This implementation is based on the algorithm presented in \cite{Knyazev2002}, which is has improved accuracy with respect to Eq.~\eqref{eq:svdangles} in the estimation of small angles.

In Fig.~\ref{fig:L96_cdv_theta}, we study the PDFs of the three principal angles between the linear subspaces for CdV and L96. In  CdV, the unstable and neutral subspaces are spanned only by the corresponding  CLVs, while the stable subspace is spanned by the remaining four CLVs, of which $\lambda_i<0$. In L96 with $D=20$, the unstable subspace is spanned by the first six CLVs, the neutral subspace is spanned only by the $7^{\text{th}}$ CLV, and the stable subspace is spanned by the remaining  13 CLVs. Focusing on Figs.~\ref{fig:L96_cdv_theta}(a-c) for CdV, we notice that this system is non-hyperbolic because the PDFs are populated close to $\theta\to0$. 
Specifically for Fig.~\ref{fig:L96_cdv_theta}(a), the binning is geometrically spaced and denser close to $\theta\to0$. Interestingly, the PDF of $\theta_{U,N}$ of CdV for small angles follows a power-law $PDF(\theta) \sim \theta^{-\alpha} $ for $\theta\to0$ and until $\approx10^\circ$, before it saturates. A different shape that is still highly non-hyperbolic is shown for the PDFs of $\theta_{U,S}$ and $\theta_{N,S}$ in Figs.~\ref{fig:L96_cdv_theta}(b-c), in which the binning is linear and both axes in logarithmic scale.  Figures~\ref{fig:L96_cdv_theta}(d-f) show the same statistics in the case of L96, which is also non-hyperbolic, as there is strong frequency of tangencies, $\theta\to0$. In all plots of Fig.~\ref{fig:L96_cdv_theta} the agreement of the subspace angle statistics between target and ESN is good, which demonstrates that the ESN has achieved a robust and accurate learning of the ergodic properties from higher-dimensional data.

Fourth, the statistics of FTCLEs ($\Lambda^c_i$), for a time-lapse of $\Delta t = m\, dt$ timesteps, in the cases of CdV and L96 are shown in Fig.~\ref{fig:L96_pdf_icle}. All six $\Lambda^c_i$ are shown for CdV, while a representative set of six $\Lambda^c_i$ are shown for L96, such that $\lambda_i>0$ for $k=1,2,3$, $\lambda_i=0$ for $k=7$, and $\lambda_i<0$ for $k=11, 17$. For CdV, the most probable statistics are  well captured by the ESN, which is in agreement with the target data. There are slight deviations at the tails of the distributions, which are still in agreement within error bars (shaded region). In the case of L96, the agreement is good for both the most probable statistics and the tails, for all FTCLEs (also those not shown). The first moment of the distributions, i.e. the mean of the FTCLEs timeseries, must be equal to the Lyapunov exponents, $\lambda_i=\frac{1}{T}\int_{0}^{T}\Lambda^c_i$, which indeed holds for all the cases considered here. The agreement between ESN and target sets in Fig.~\ref{fig:L96_pdf_icle} shows that the ESN is able to accurately learn the finite-time variability of the CLV growth rates also for higher dimensional systems that are characterized by many Lyapunov exponents.

Finally, in Table~\ref{tab:KYdim} we show the estimated Kaplan-Yorke dimension \cite{KaplanYorke1979} for all the considered systems and compare the outcomes of the ESN and target. This dimension is an upper bound of the attractor's fractal dimension\cite{Ott2002}, which is defined as
\begin{equation}
D_{\text{KY}} = k + \frac{\sum_{i=1}^{k} \lambda_{i}}{|\lambda_{i+1}|},
\label{eq:DKY}
\end{equation}
where $k$ is such that the sum of the first $k$ LEs is positive and the sum of the first $k+1$ LEs is negative. We observe a  good agreement in all cases with  $\leq1\%$ error. This observation further confirms the ability of the ESN to accurately learn the properties of the chaotic attractor.

\begin{table}[]
\caption{Estimates of the Kaplan-Yorke dimension for all attractors, comparing between the target and echo state network. The error is the quantity $\frac{\rm target-ESN}{\rm target}\times100\%$.}
\renewcommand{\arraystretch}{1.2}
\centering
\begin{tabular}{lccl}
	\hline\noalign{\smallskip}
	\multicolumn{1}{c}{} & Target & ESN    & \% error \\
	\noalign{\smallskip}\hline\noalign{\smallskip}
	Lorenz 63            & 2.0621 & 2.0618 & 0.015    \\
	Rossler              & 2.0051 & 2.0049 & 0.01    \\
	CdV                  & 2.294  & 2.277  & 0.74    \\
	Lorenz 96            & 13.4697 & 13.4721 & 0.018   \\ 
	\noalign{\smallskip}\hline
\end{tabular}
\label{tab:KYdim}
\end{table}

\section{Conclusions}\label{sec:conc}

Stability analysis is a principled mathematical tool to quantitatively answer key questions on the behavior of nonlinear systems: 
Will infinitesimal perturbations grow in time (i.e., is the system linearly unstable)? 
If so, what are the perturbations' growth rates (i.e.,  how linearly unstable is the system)? 
What are the directions of growth?
To answer these questions, traditionally, we linearize the equations of the dynamical system around a reference point,  and compute the properties of the tangent space, the dynamics of which is governed by the Jacobian. The overarching goal of this paper is to propose a method that infers the stability properties directly from data, which does not rely on the knowledge of the dynamical differential equations.
We tackle chaotic systems, which have a linearized behavior that is more general and intricate than periodic or quasi-periodic oscillations.  
First, we propose the Echo State Network with the Recycle Validation as a tool to accurately learn the chaotic dynamics from data. The data is provided by the integration of low- and higher- dimensional prototypical chaotic dynamical systems. These systems are qualitatively  different from each other, and are toy models that describe diverse physical settings, ranging from climatology and meteorology to chemistry.  
Second, we mathematically derive the Jacobian of the Echo State Network (Eq.~\eqref{eq:genesnJac}). 
In contrast to other recurrent neural networks, such as long short-term memory networks or gated recurrent units, the Jacobian of the ESN is mathematically simple and computationally straightforward.
Third, we analyse the stability properties inferred from the ESN, and compare them with the target properties (ground truth) obtained by linearizing the equations. The ESN correctly infers quantities that characterize the chaotic dynamics and its tangent space 
(i) the long-term statistics of the solution, for which we compute the probability density function of each state variable;
(ii) the covariant Lyapunov vectors, which are a physical basis for the tangent space that is covariant with the dynamics; 
(iii) the Lyapunov spectrum, which is the set of eigenvalues of the Oseledets matrix that are the perturbations' average exponential growths;
(iv) the finite-time Lyapunov exponents, which are the finite-time growth along the covariant Lyapunov vectors; and 
(v) the angles between the stable, neutral, and unstable splittings of the tangent space, which informs about the degree of hyperbolicity of the attractor. We show that these quantities can be accurately learned from data by the ESN, with negligible numerical errors.

As mathematically and numerically shown in~\cite{Huhn_Magri_2020}, the stability properties of fixed points (with eigenvalue analysis) and periodic solutions (with Floquet analysis) can be inferred from covariant Lyapunov analysis. Therefore, this work opens up new opportunities for the inference of stability properties from data in nonlinear systems, from simple fixed points, through periodic oscillations, to chaos. 

\begin{acknowledgements}
This research has received financial support from the ERC Starting Grant No. PhyCo 949388. LM gratefully acknowledges financial support TUM Institute for Advanced Study (German Excellence Initiative and the EU 7th Framework Programme No. 291763). We are grateful to Alberto Racca for insightful discussions regarding the ESN. GM is also grateful to Valerio Lucarini for insightful discussions regarding dynamical systems theory.
\end{acknowledgements}

\section*{Data availability}
The implementation of the ESN follows\cite{Racca2021} and the code can be found in the github repository \href{https://github.com/gmargazo/ESN-CLVs.git}{https://github.com/gmargazo/ESN-CLVs.git}. 

\section*{Conflict of interest}

The authors declare that they have no conflict of interest.

\appendix

\section{Algorithms to compute LEs and CLVs}\label{app:1}

In this section, we present two algorithms for the computation of LEs and CLVs. Algorithm \ref{alg:LEs} is used to calculate the first $D$ LEs of an ESN, where $N_r$ is the dimensionality of the hidden state and $D$ is the dimensionality of the input state. This algorithm follows the methods described in \cite{Pathakchaos2017,Vlachas2020}. Algorithm \ref{alg:CLVs} computes the first $D$ CLVs for both the ESN and target chaotic systems, using the approach outlined in \cite{Ginelli2007}. These algorithms are crucial for understanding the dynamics and predictability of the systems being studied.

\RestyleAlgo{ruled}

\SetCommentSty{footnotesize}
\SetKwComment{Comment}{/* }{ */}

\begin{algorithm}[hbt!]
\caption{Algorithm to calculate the Lyapunov exponents of the echo state network}\label{alg:LEs}
$\bU \gets \texttt{random}\in\mathbb{R}^{N_r\times D}$ \Comment*[r]{Initialize $D$ GSVs}
$\bQ, \bR \gets \text{QR}(\bU)$  \Comment*[r]{Othonormalize GSVs}
$\bU \gets \bQ\in\mathbb{R}^{N_r\times D}$\;
$N^{QR} \gets (N^{test}-N^w)/m$ \Comment*[r]{Number of QR decompositions}	
\textit{Save the timeseries of $\bR$ and $\bQ$ for CLVs calculation}\\
Initialize $\tilde{\bR} \gets \boldsymbol{0}\in\mathbb{R}^{D\times D\times N^{QR}}$\;
Initialize $\tilde{\bQ} \gets \boldsymbol{0}\in\mathbb{R}^{N_r\times D\times N^{QR}}$\;
Initialize $\Lambda \gets \boldsymbol{0}\in\mathbb{R}^{D\times N^{QR}}$ \Comment*[r]{Save the FTLEs}
$\bW^c= \mathbf{W}^T_{\mathrm{in}}\mathbf{W}^T_{\mathrm{out}} + \mathbf{W}^T$  \Comment*[r]{Constant matrices of Eq.~\eqref{eq:genesnJac}}

\textit{Evolve the hidden state and GSVs simultaneously.}\\
\textit{Skip a transient initial $N^w$ steps for warm-up.}\\
$n\gets0$ \Comment*[r]{Increments the number of QR decompositions}
\For{$i=0:N^{test}$}{ 
	$\br(\rt_{i+1})\gets  \tanh\left([\hat{\by}(\rt_{i});b_\mathrm{in}]^T\mathbf{W}_{\mathrm{in}}+\br(\rt_{i})^T\mathbf{W}\right)$\;	
	$\by(\rt_{i+1}) \gets [\br(\rt_{i+1});1]^T \textbf{W}_{\mathrm{out}}$\;		
	$\bJ \gets (1 - \br(\rt_{i+1})^2)\bW^c$ \Comment*[r]{The updated Jacobian}
	$\bU \gets \bJ \bU$ \Comment*[r]{The variational equation}
	\If{$\mod(i,m)=0$}{
		$\bQ, \bR \gets \text{QR}(\bU)$  \Comment*[r]{QR every $m$ steps}
		$\bU \gets \bQ$\;
		\If{$i>N^w$}{
			$\Lambda[:,n] \gets \log{\left( \text{diag}[\bR] \right)}/dt$ \Comment*[r]{Save the FTLEs}		
			$\tilde{\bR}[:,:,n] \gets \bR$ \Comment*[r]{Save $\bR$}
			$\tilde{\bQ}[:,:,n] \gets \bQ$ \Comment*[r]{Save $\bQ$}
			$n \gets n +1$\;
		}
	}
	
}
$\lambda_j \gets \sum_{i=0}^{N^{QR}}{\Lambda[j,i]}/T^{test}$ \Comment*[r]{The \textit{j}th Lyapunov exponent}
\end{algorithm}

\begin{algorithm}[hbt!]
\caption{Algorithm to calculate the covariant Lyapunov vectors\cite{Ginelli2007,Ginelli2013}}\label{alg:CLVs}
\KwData{Given $\tilde{\bR}$ and $\tilde{\bQ}$}
\textit{Set to $\mathbf{0}$ the matrices $\bC$, $\bD$ and $\bV$}\\
$\bC \gets \boldsymbol{0}\in\mathbb{R}^{D\times D\times N^{QR}}$ \Comment*[r]{Coordinates of CLVs in the GSV basis}
$\bD \gets \boldsymbol{0}\in\mathbb{R}^{D\times D\times N^{QR}}$ \Comment*[r]{Growth factors of CLVs}
$\Lambda^c \gets \boldsymbol{0}\in\mathbb{R}^{D\times N^{QR}}$ \Comment*[r]{The $D$ FTCLEs}
$\bV \gets \boldsymbol{0}\in\mathbb{R}^{N_r\times D\times N^{QR}}$ \Comment*[r]{Each column is a CLV}
\textit{Set final time index to identity $\mathbb{I}$ for the matrices $\bC$ and $\bD$}\\
$\bC[:,:,N^{QR}] \gets \mathbb{I}$\;
$\bD[:,:,N^{QR}] \gets \mathbb{I}$\;
$\bV[:,:,N^{QR}] \gets \tilde{\bQ}[:,:,N^{QR}] \bC[:,:,N^{QR}]$\;

\textit{Evolve backwards and solve Eq.~\eqref{eq:C}}\\

\For{$i=N^{QR}-1:0$}{ 
	$\bG \gets \texttt{solve\_triangular}(\tilde{\bR}[:,:,i]\bC[:,:,i+1])$\;
	\textit{Normalize each column of $\bG$}\\
	\For{$j=0:\bG.shape[1]$}{
		$\bD[:,:,i] \gets \texttt{norm}(\bG[:,j])$\;
		$\bC[:,:,i] \gets \bG[:,j]/\bD[:,:,i]$\;}
	
	$\bV[:,:,i] \gets \tilde{\bQ}[:,:,i] \bC[:,:,i]$\Comment*[r]{Calculate CLVs}
	$\Lambda^c[:,i] \gets \log{\left( \text{diag}[\bD[:,:,i]] \right)}/dt$ \Comment*[r]{Calculate FTCLEs}
	
}

\end{algorithm}

\section{Robustness}\label{app:2}
An important aspect of data-driven approaches is their ability to perform accurately under a variety of conditions.  In this section, we evaluate the robustness of our approach by using smaller training sets (less data) subject to  noise levels that are higher than those of  Sect.~\ref{sec:results}. We also test the effect of using a loss function other than the mean square error (MSE), as defined in Eq.~\eqref{eq:MSE}, on the accuracy of the learning. The ESN architecture follows~\cite{Racca2021}, where it was trained with chaotic data from the Lorenz 63 and Lorenz 96 systems, and was robustly optimized to maximize the prediction horizon under different validation strategies. 

\subsection{Training with less data and higher noise intensity}\label{app:2noisy}

\begin{figure}[t]
	\centering
	\includegraphics[width=.4\textwidth]{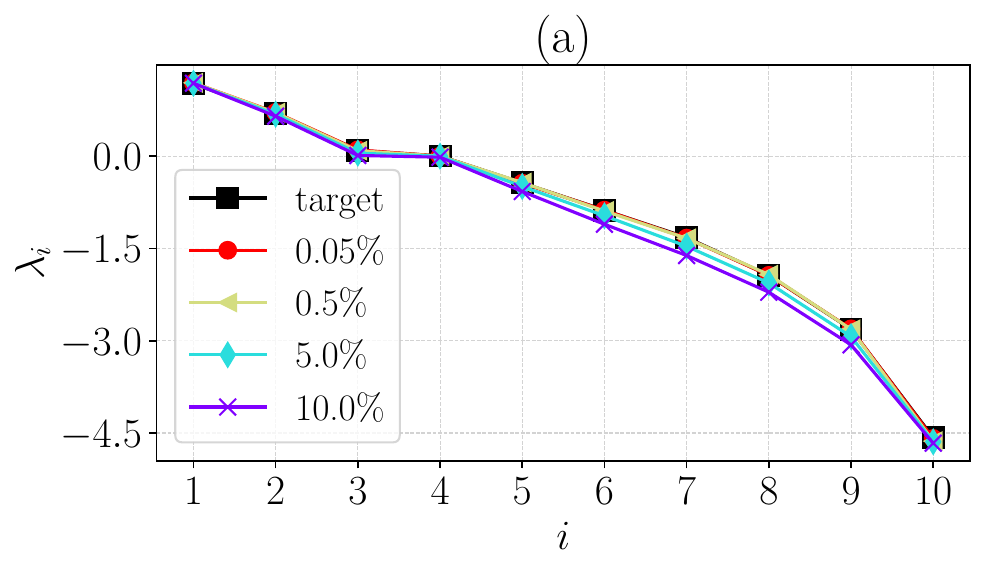}
	\includegraphics[width=.4\textwidth]{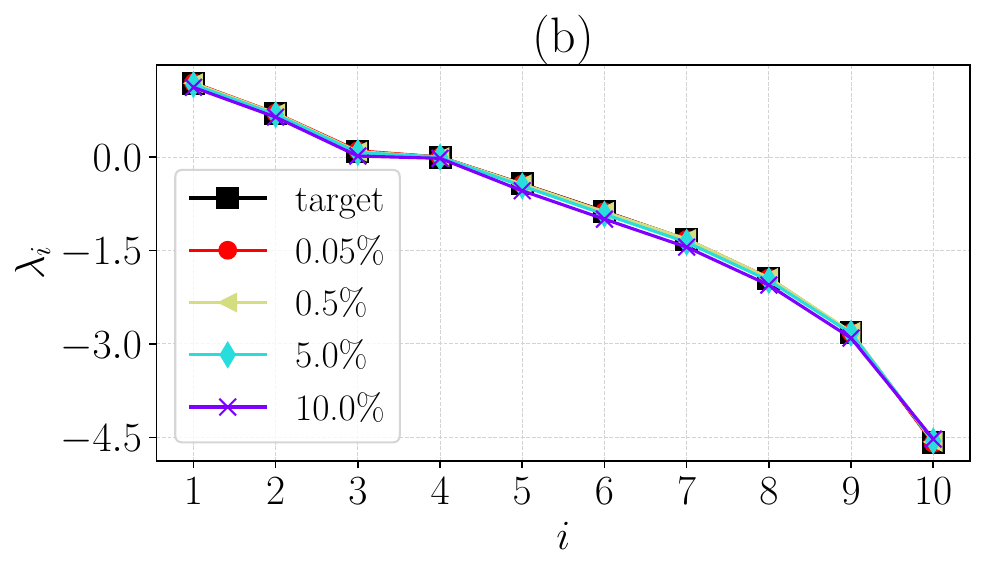}
	\caption{Lyapunov spectrum of Lorenz 96 trained with (a) $100\tau_\lambda$ and (b) $500\tau_\lambda$ long time series, and different noise intensity, as indicated in the legend.}
	\label{fig:L96_LEs_ntrain}
\end{figure}

It has been demonstrated that adding a small amount of Gaussian centered noise proportional to the standard deviation of the chaotic signal during training can improve the performance of an ESN~\cite{Vlachas2020,Racca2021}. Noise aids the ESN to generalize  to unseen data. In Sect.~\ref{sec:results} we add Gaussian noise with a zero mean and standard deviation, $\sigma_n=\delta\sigma_y$, where $\delta=0.05\%$, and $\sigma_y$ is the standard deviation of the data component-wise. We consider the Lorenz 96 with $D=10$ degrees of freedom and $F=8$, such that the system is chaotic. We increase the noise intensity to $\delta=\{0.5\%, 5\%, 10\%\}$. We also quantify the effect of less training data by using $100\tau_\lambda$ and $500\tau_\lambda$ long time series, i.e.~$1/10$ and half of the $1000\tau_\lambda$ long time series that we used in Sect.~\ref{sec:results}. Figure \ref{fig:L96_LEs_ntrain} shows the effects in the Lyapunov spectrum. For \ref{fig:L96_LEs_ntrain}(a), where the training set is $100\tau_\lambda$ long, there is a good agreement between the target (black squares) and the ESN (colored points) positive exponents. As expected, a gradual deterioration appears as the noise increases. In \ref{fig:L96_LEs_ntrain}(b) for a  $500\tau_\lambda$ long training set, the agreement is good for all exponents with a smaller difference for negative exponents compared to (a). 
After training $N_{\mathrm{ESN}}=10$ statistically independent networks with chaotic timeseries, some might eventually evolve towards a fixed point or a periodic orbit instead (i.e.~they show spurious behavior). Here, for $100\tau_\lambda$ long training timeseries, no ESN evolves spuriously at 0.05\% and 0.5\% noise. However, at 10\% noise, half of the networks show spurious evolution, and are discarded at postprocessing. Instead, for $500\tau_\lambda$ long training timeseries, one and two out of ten evolves spuriously at 0.05\% and 0.5\% noise, respectively, but none at 5\% and 10\% noise, which ensures robustness of the network. 

\begin{figure}[t]
	\centering
	\includegraphics[width=.5\textwidth]{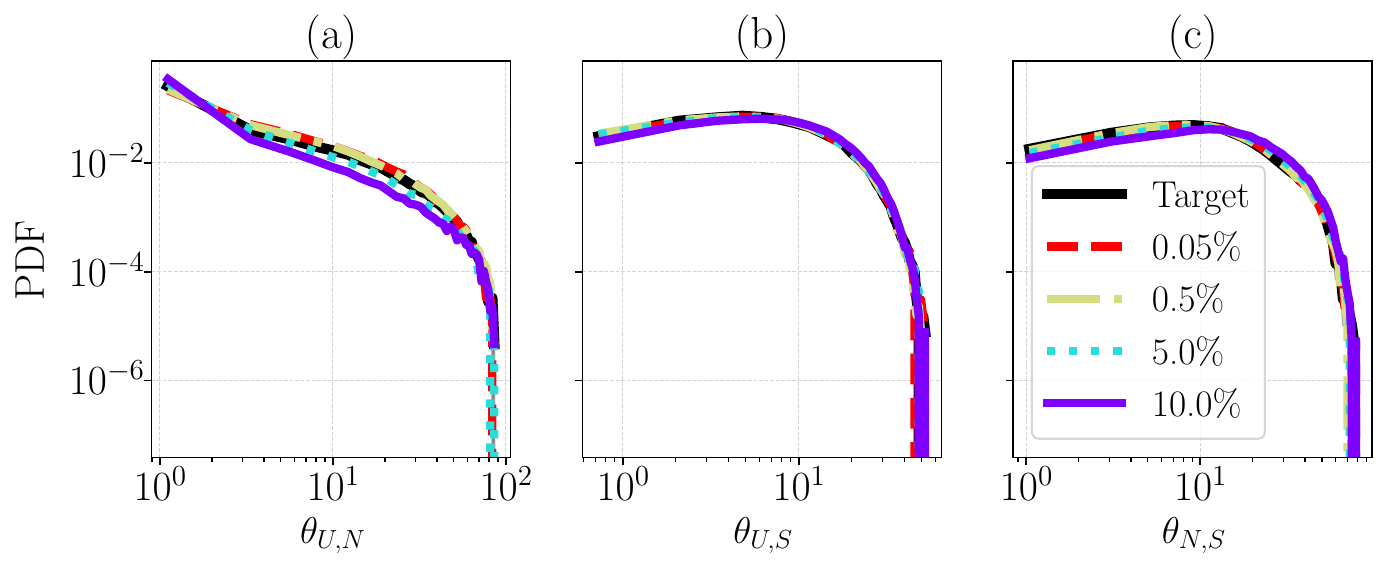}
	\includegraphics[width=.5\textwidth]{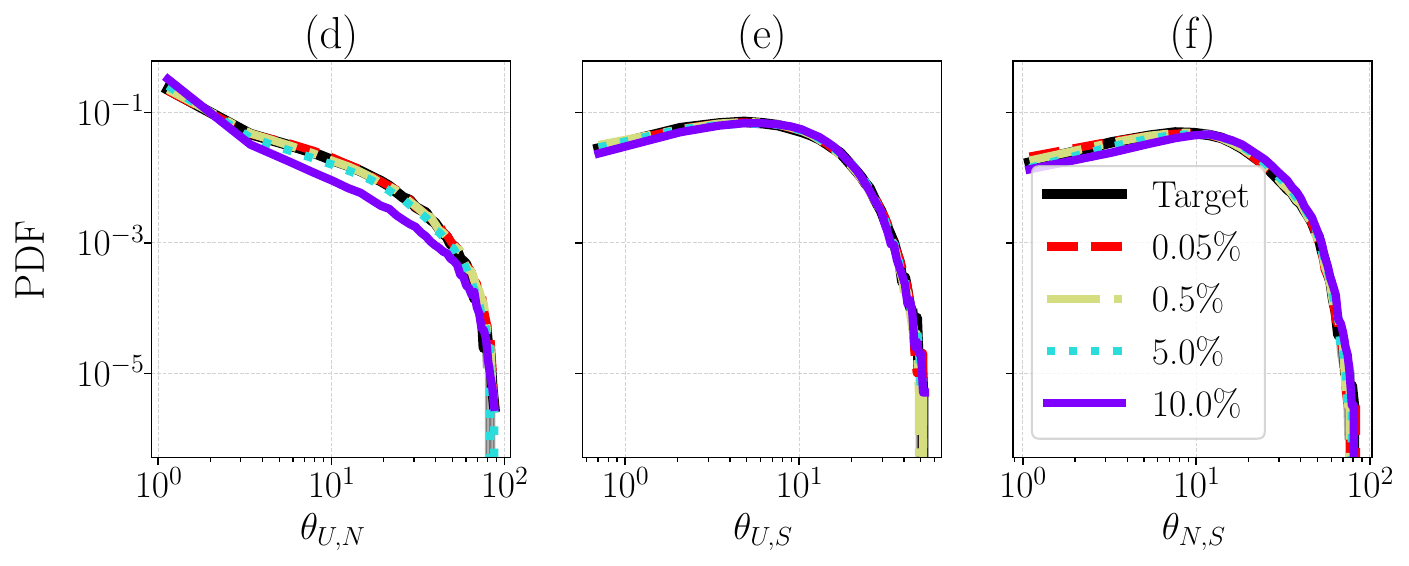}
	\caption{PDF of minimum angles between subspaces of CLV from Lorenz 96 trained with (a-c) $100\tau_\lambda$ and (d-f) $500\tau_\lambda$ long time series, and different noise intensity, as indicated in the legend. Both $x$ and $y$ axes are in logarithmic scale and the $x$-axis is in degrees.}
	\label{fig:L96_clvs_ntrain}
\end{figure}

As a further test, in Fig.~\ref{fig:L96_clvs_ntrain} we consider the minimum angles between subspaces spanned by CLVs. In \ref{fig:L96_clvs_ntrain}(a-c) the ESNs are trained with $100\tau_\lambda$ long timeseries, and accordingly in \ref{fig:L96_clvs_ntrain}(d-f) with $500\tau_\lambda$. Overall, the results are in good agreement with the target ensuring the robustness of the ESN. A slight and gradual disagreement is observed as the noise intensity increases, in particular for $\theta_{U,N}$.

\subsection{Training with a different loss function}\label{app:2loss}

\begin{figure}[t]
	\centering
	\includegraphics[width=.4\textwidth]{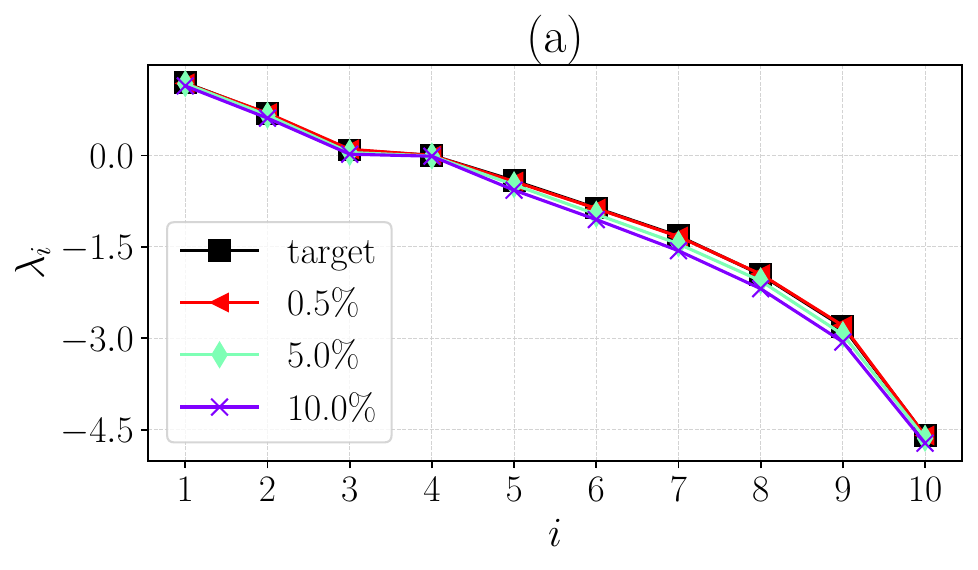}
	\includegraphics[width=.5\textwidth]{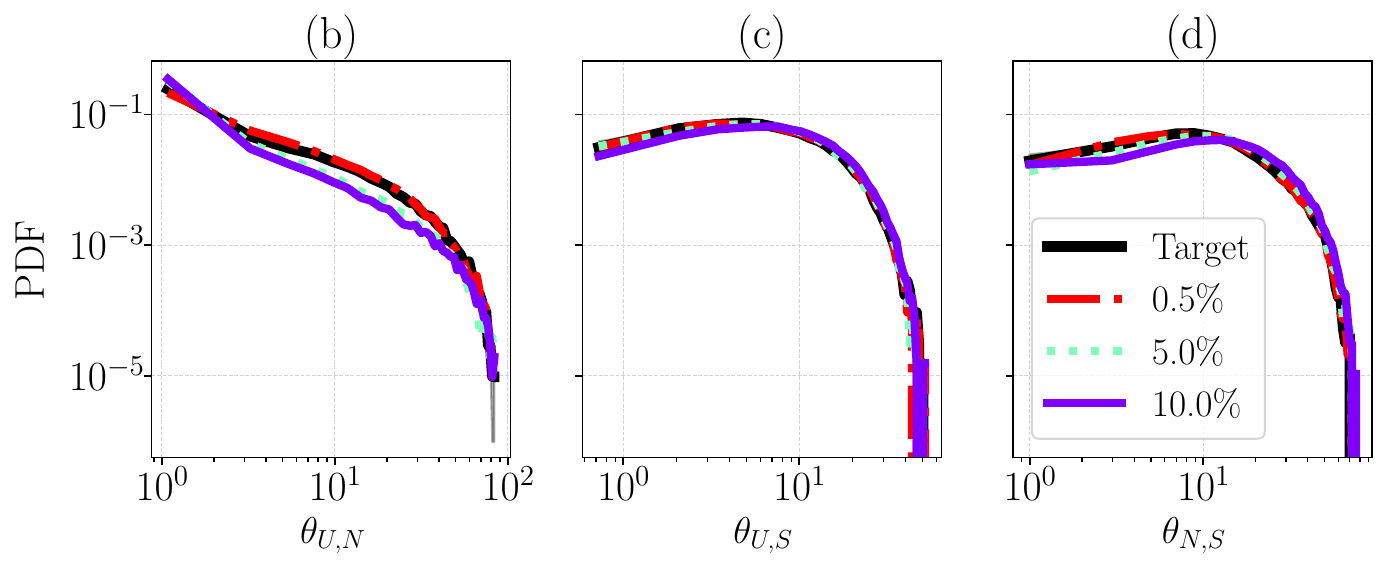}
	\caption{Using the mean absolute error, Eq.~\eqref{eq:MAE}, to train the ESN with  $100\tau_\lambda$ long time series from the Lorenz 96, and with different noise intensity, as indicated in the legends. (a) Lyapunov spectrum. (b-d) PDF of minimum angles between subspaces of CLVs, where both $x$ and $y$ axes are in logarithmic scale and the $x$-axis is in degrees.}
	\label{fig:L96_loss}
\end{figure}

The mean square error (MSE), Eq.\eqref{eq:MSE}, is a commonly used loss function in the ESN architecture\cite{Lukosevicius2012}. We investigate the effect of using a mean absolute error (MAE) loss function defined as
\begin{equation}
	\textrm{MAE} = \frac{1}{N_{\mathrm{tr}}N_y} \sum_{i=0}^{N_{\mathrm{tr}}} | \by_{\mathrm{p}}(\rt_{i}) - \by_{\mathrm{in}}(\rt_{i})|.
	\label{eq:MAE}
\end{equation}
By comparing the stability properties obtained using the MSE and MAE loss functions, we can gain a better understanding of the potential impact of the choice of loss function on the performance of ESN. In Fig.~\ref{fig:L96_loss} the results correspond to a $100\tau_\lambda$ long training set, where Eq.~\ref{eq:MAE} was used as a loss function. The Lyapunov spectrum of Fig.~\ref{fig:L96_loss}(a) is qualitatively similar to Fig.~\ref{fig:L96_LEs_ntrain}(a). In practice, training with MAE resulted in less stable ESNs, with increased failures during the test set. For a $100\tau_\lambda$ long training set, at $10\%\sigma_y$ noise with MAE, 80\% of ESNs failed, in contrast to 50\% with MSE for the same noise. Figures ~\ref{fig:L96_loss}(b-d) are similar to Figs.~\ref{fig:L96_clvs_ntrain}(a-c) showing minor differences. We also trained the ESNs with $500\tau_\lambda$ long training sets, as in Sect.~\ref{app:2noisy}. Interestingly, we obtain similar results with Fig.~\ref{fig:L96_LEs_ntrain}(b) and Figs.~\ref{fig:L96_clvs_ntrain}(d-f), with no significant differences (result not shown).

Based on our analyses, we can conclude that the process of extracting the stability properties of an ESN is robust against higher levels of noise, smaller training sets, and the use of a MAE loss function. Our results suggest that a good practice is to use small to moderate levels of centered Gaussian noise in the training set, a sufficiently large reservoir size, and a training trajectory of at least $100\tau_\lambda$. 

\bibliographystyle{spphys}       
\bibliography{bibliography.bib}

\end{document}